\documentclass[twocolumn]{aastex631}

\usepackage{amsmath}	
\usepackage{amssymb}	
\usepackage{booktabs}
\usepackage{lipsum}
\usepackage[T1]{fontenc}
\usepackage{graphicx}	

\usepackage{newtxtext,newtxmath}
\usepackage{orcidlink}
\usepackage{pifont}
\usepackage{natbib}
\usepackage{xparse}
\usepackage{ifthen}
\usepackage{overpic}

\usepackage{savesym}
\savesymbol{tablenum}
\usepackage{siunitx}
\usepackage{tikz}
\restoresymbol{SIX}{tablenum}

\newcommand{\Gaia}{\textit{Gaia}}

\newcommand{\WISE}{\textit{WISE}}
\newcommand{\GALEX}{\textit{GALEX}}

\newcommand{\xposa}{12}
\newcommand{\yposa}{4}
\newcommand{\xposb}{15}
\newcommand{\yposb}{52}
\newcommand{\xposc}{14}
\newcommand{\yposc}{12}
\newcommand{\xposd}{14}
\newcommand{\yposd}{12}

\newcommand{\customfigure}[2]{
\begin{figure*}[t]
    \centering
    \hspace*{-0.025\linewidth}
    \begin{tabular}{@{}c@{\hskip 0.0001\linewidth}c@{}}
    \makebox[0.5\linewidth][l]{\hspace{0.5cm}
        \begin{overpic}[width=0.5\linewidth]{anc/#1/sharkvis.jpeg}
            \put(\xposa, \yposa){\Large\textcolor{white}{(a)}}
        \end{overpic}} & 
    \begin{overpic}[width=0.5\linewidth]{anc/#1/rv_orbit.pdf}
        \put(\xposb, \yposb){\Large(b)}
    \end{overpic} \\
    \begin{overpic}[width=0.5\linewidth]{anc/#1/companion_mass.pdf}
        \put(\xposc, \yposc){\Large(c)}
    \end{overpic} &
    \begin{overpic}[width=0.5\linewidth]{anc/#1/sed_fit.pdf}
        \put(\xposd, \yposd){\Large(d)}
    \end{overpic} \\
    \end{tabular}
    \caption{#2}
    \label{fig:panel_#1}
\end{figure*}
}

\newcommand{\xposaa}{12}
\newcommand{\yposaa}{4}
\newcommand{\xposbb}{14}
\newcommand{\yposbb}{10.5}
\newcommand{\xposcc}{14}
\newcommand{\yposcc}{12}
\newcommand{\xposdd}{14}
\newcommand{\yposdd}{12}
\newcommand{\xposee}{8.5}
\newcommand{\yposee}{6.5}

\newcommand{\customfigureilocator}[2]{
\begin{figure*}[t]
    \centering
    \hspace*{-0.025\linewidth}
    \begin{tabular}{@{}c@{\hskip 0.0001\linewidth}c@{}}
    \makebox[0.5\linewidth][l]{\hspace{0.35cm}
        \begin{overpic}[width=0.5\linewidth]{anc/#1/sharkvis.jpeg} 
            \put(\xposaa, \yposaa){\Large\textcolor{white}{(a)}}
        \end{overpic}} &  
    \begin{overpic}[width=0.5\linewidth]{anc/#1/rv_orbit.pdf} 
        \put(\xposbb, \yposbb){\Large(b)}
    \end{overpic} \\
    \begin{overpic}[width=0.5\linewidth]{anc/#1/companion_mass.pdf} 
        \put(\xposcc, \yposcc){\Large(c)}
    \end{overpic} & 
    \begin{overpic}[width=0.5\linewidth]{anc/#1/sed_fit.pdf} 
        \put(\xposdd, \yposdd){\Large(d)}
    \end{overpic} \\
    \end{tabular}
    
    \vspace{0.05cm} 
    \begin{overpic}[width=\linewidth]{anc/#1/ilocator.pdf}
        \put(\xposee, \yposee){\Large(e)}
    \end{overpic}
    \caption{#2}
    \label{fig:panel_#1}
\end{figure*}
}

\begin{document}

\title{Hidden in Plain Sight: Searching for Dark Companions to Bright Stars \\with the Large Binocular Telescope and SHARK-VIS}

\correspondingauthor{D. M. Rowan}
\email{rowan.90@osu.edu}

\author[0000-0003-2431-981X]{D. M. Rowan}
\affiliation{Department of Astronomy, The Ohio State University, 140 West 18th Avenue, Columbus, OH, 43210, USA}
\affiliation{Center for Cosmology and Astroparticle Physics, The Ohio State University, 191 W. Woodruff Avenue, Columbus, OH, 43210, USA}

\author[0000-0003-2377-9574]{Todd A. Thompson}
\affiliation{Department of Astronomy, The Ohio State University, 140 West 18th Avenue, Columbus, OH, 43210, USA}
\affiliation{Center for Cosmology and Astroparticle Physics, The Ohio State University, 191 W. Woodruff Avenue, Columbus, OH, 43210, USA}
\affiliation{Department of Physics, The Ohio State University, Columbus, Ohio, 43210, USA}

\author{C. S. Kochanek}
\affiliation{Department of Astronomy, The Ohio State University, 140 West 18th Avenue, Columbus, OH, 43210, USA}
\affiliation{Center for Cosmology and Astroparticle Physics, The Ohio State University, 191 W. Woodruff Avenue, Columbus, OH, 43210, USA}

\author[0000-0001-9539-2112]{G. Li Causi}
\affiliation{INAF-National Institute for Astrophysics, Osservatorio Astronomico di Roma, Via Frascati 33, \\I-00078 Monte Porizo Catone, Rome, Italy}
\affiliation{INAF-ADONI, Adaptive Optics National Laboratory, Rome, Italy}

\author[0000-0002-8873-8215]{J. Roth}
\affiliation{Max-Planck-Institut für Astrophysik, Karl-Schwarzschild-Str. 1, 85748 Garching, Germany}
\affiliation{Technical University of Munich, Boltzmannstr. 3, 85748 Garching, Germany}

\author[0009-0003-7811-1683]{P. Vaccari}
\affiliation{INAF-National Institute for Astrophysics, Osservatorio Astronomico di Roma, Via Frascati 33, \\I-00078 Monte Porizo Catone, Rome, Italy}
\affiliation{INAF-ADONI, Adaptive Optics National Laboratory, Rome, Italy}

\author[0000-0002-0983-8040]{F. Pedichini}
\affiliation{INAF-National Institute for Astrophysics, Osservatorio Astronomico di Roma, Via Frascati 33, \\I-00078 Monte Porizo Catone, Rome, Italy}
\affiliation{INAF-ADONI, Adaptive Optics National Laboratory, Rome, Italy}

\author[0000-0002-2264-8698]{R. Piazzesi}
\affiliation{INAF-National Institute for Astrophysics, Osservatorio Astronomico di Roma, Via Frascati 33, \\I-00078 Monte Porizo Catone, Rome, Italy}
\affiliation{INAF-ADONI, Adaptive Optics National Laboratory, Rome, Italy}

\author[0000-0002-0666-3847]{S. Antoniucci}
\affiliation{INAF-National Institute for Astrophysics, Osservatorio Astronomico di Roma, Via Frascati 33, \\I-00078 Monte Porizo Catone, Rome, Italy}
\affiliation{INAF-ADONI, Adaptive Optics National Laboratory, Rome, Italy}

\author[0000-0003-1033-1340]{V. Testa}
\affiliation{INAF-National Institute for Astrophysics, Osservatorio Astronomico di Roma, Via Frascati 33, \\I-00078 Monte Porizo Catone, Rome, Italy}
\affiliation{INAF-ADONI, Adaptive Optics National Laboratory, Rome, Italy}

\author[0000-0002-5099-8185]{M. C. Johnson}
\affiliation{Department of Astronomy, The Ohio State University, 140 West 18th Avenue, Columbus, OH, 43210, USA}

\author[0000-0002-1503-2852]{J. Crass}
\affiliation{Department of Astronomy, The Ohio State University, 140 West 18th Avenue, Columbus, OH, 43210, USA}
\affiliation{University of Notre Dame, Physics and Astronomy, Notre Dame, IN 46556, USA}

\author[0000-0003-0800-0593]{J. R. Crepp}
\affiliation{University of Notre Dame, Physics and Astronomy, Notre Dame, IN 46556, USA}

\author[0000-0002-3047-9599]{A. Bechter}
\affiliation{University of Notre Dame, Physics and Astronomy, Notre Dame, IN 46556, USA}

\author[0000-0001-8725-8730]{E. B. Bechter}
\affiliation{University of Notre Dame, Physics and Astronomy, Notre Dame, IN 46556, USA}

\author[0009-0000-4650-2266]{B. L. Sands}
\affiliation{University of Notre Dame, Physics and Astronomy, Notre Dame, IN 46556, USA}

\author[0000-0002-8518-4640]{R. J. Harris}
\affiliation{Ctr. for Advanced Instrumentation, Durham Univ., Durham, United Kingdom}



\begin{abstract}

We report the results from a pilot study to search for black holes and other dark companions in binary systems using direct imaging with SHARK-VIS and the iLocater pathfinder ``Lili'' on the Large Binocular Telescope. Starting from known single-lined spectroscopic binaries, we select systems with high mass functions that could host dark companions and whose spectroscopic orbits indicate a projected orbital separation $\geq 30$~mas. For this first exploration, we selected four systems (HD~137909, HD~104438, HD~117044, and HD~176695). In each case, we identify a luminous companion and measure the flux ratio and angular separation. However, two of the systems (HD~104438 and HD~176695) are not consistent with simple binary systems and are most likely hierarchical triples. The observed companions rule out a massive compact object for HD~137909, HD~117044, and HD~176695. HD~104438 requires further study because the identified star cannot be responsible for the RV orbit and is likely a dwarf tertiary companion. The SHARK-VIS observation was taken near pericenter, and a second image near apocenter is needed to discriminate between a closely separated luminous secondary and a compact object. We show how the combination of RVs and direct imaging can be used to constrain the orbital inclination and companion mass, and discuss the potential of high resolution direct imaging surveys to identify and confirm non-interacting compact object candidates.

\end{abstract}

\keywords{Binary stars(154) ---  Compact objects(288) --- Direct imaging(387)}


\section{Introduction} \label{sec:intro}

The mass distribution of compact objects is closely linked to the evolution of massive stars and supernovae. The final fate of massive stars depends on a number of factors including the progenitor star's composition, mass-loss rates, and past binary interactions \citep[e.g.,][]{Sukhbold16}. There are estimated to be $\sim 10^8$ stellar mass black holes (BHs) and $\sim 10^9$ neutron stars (NSs) in the Milky Way \citep{Timmes96}, but only a tiny fraction have been observed. 

The majority of the observed compact objects are in X-ray binaries \citep[e.g.][]{MataSanchez24} or were discovered in gravitational wave mergers \citep[e.g.,][]{Abbott16}. There are expected to be only a few thousand X-ray binaries in the Milky Way, since most must be in tight orbits to produce observable X-ray emission \citep{CorralSantana16}. The gravitational wave merger sources are all extragalactic, and detections are biased towards more massive BHs \citep[$M \sim 20$--$30\ M_\odot$,][]{Abbott19}, although systems with lower mass components have also been identified \citep{Abbott23}. Only a tiny fraction of the BHs and NSs in the Milky Way are expected to evolve into gravitational wave merger sources. 

Instead, the vast majority of BHs and NSs are expected to be either free-floating, isolated systems, or in wide, non-interacting binaries. Isolated BHs can only be detected via microlensing surveys \citep[e.g.,][]{Lam22, Lam23, Sahu22}, and the upcoming Roman and Rubin time-domain surveys could detect tens to hundreds of isolated black hole microlensing events, depending on the final survey configurations \citep{Abrams23, Lam23_whitepaper}. 

There are multiple observational methods that can be used to search for compact objects in binary systems. The third \Gaia{} data release \citep[\Gaia{} DR3,][]{GaiaDR3} includes $>$130,000 astrometric orbits. By combining the \Gaia{} astrometric orbit and radial velocities from \Gaia{} and other ground-based facilities, three BHs have been confidently detected. \Gaia{}-BH1 \citep{ElBadry23_BH1, Chakrabarti23} is binary with a $9.6\ M_\odot$ BH and a Sun-like star and an orbital period of 185~days. The system is at $480$~pc, making it the nearest known BH. \Gaia{}-BH2 \citep{Tanikawa23, ElBadry23_BH2} is a binary with a $8.9\ M_\odot$ BH and a red giant in a longer 1277~day orbit at a distance of 1.16~kpc. Standard binary evolution formation channels struggle to explain the observed orbits of these systems, and alternative formation scenarios from dynamical interactions in clusters have been proposed \citep{Tanikawa24}. Finally, \Gaia{}-BH3 is much more massive BH with $M=33\ M_\odot$ in a 11.6~yr binary with a metal-poor giant in the MW halo at a distance of $590$~pc \citep{GaiaBH3}. This system could have formed as a primordial binary \citep[e.g.,][]{ElBadry24_BH3} or through dynamical interactions in a stellar stream \citep[e.g.,][]{MarinPina24}. The three \Gaia{} BHs are all closer than the majority of X-ray binaries ($\sim 2$--10~kpc), suggesting many more systems must exist. \Gaia{} astrometry has also been used to detect 21 NS candidates \citep{ElBadry24_NS}. The systems span a wide range of orbital periods ($P\sim 180$--$1000$~days) and distances ($d\sim 240$--$1600$~kpc).

While future \Gaia{} data releases are expected to continue to expand the population of non-interacting compact object binaries \citep[e.g.,][]{Breivik17}, there are other avenues to detect and characterize systems with dark companions. For short period systems, photometric variability from ellipsoidal variations can be used to identify stars gravitationally distorted by massive companions \citep[e.g.,][]{Rowan21, Gomel21, Green23}. The orbital inclination of the system can be measured by modeling the ellipsoidal modulations, so a companion mass can be precisely determined given an estimate of the photometric primary mass. However, this method is prone to false positives from contact binaries \citep[e.g.,][]{Nagarajan23, ODoherty23} or stripped-star imposters \citep[e.g.,][]{Thomspon19, Jayasinghe22, ElBadry22_zoo, Rowan24}, and no unambiguous non-interacting BHs or NSs have been identified starting from photometric surveys. 

Spectroscopic surveys can also be used to identify stars with large amplitude, periodic radial velocity variability. For example, \citet{Giesers18} identified a stellar mass black hole in NGC~3201 with mass $M\gtrsim4.4\ M_\odot$. \citet{Song24} recently identified a $3.6\ M_\odot$ BH candidate in the \Gaia{} DR3 single-lined spectroscopic binary catalog (SB1), but additional RVs are needed to confirm the orbit. There have also been a number of searches \citep[e.g.,][]{Yi22, Yuan22} for compact objects in the Large Sky Area Multi-Object Fiber Spectroscopic Telescope \citep[LAMOST,][]{Cui12}, but no strong BH candidates have been identified.

The primary limitation of spectroscopic-based searches, as compared to astrometric and photometric methods, is that the orbital inclination cannot be constrained by RVs alone. Even with an estimate of the mass of the photometric primary, only the minimum companion mass can be determined. Systems at lower inclinations with black hole companions could possibly be missed when selecting candidates based on their RV amplitude and period \citep[e.g.,][]{Jayasinghe23}.

The observational methods used to search for non-interacting black hole companions parallel those used to identify and characterize exoplanets. Thousands of exoplanet candidates have been characterized using transits \citep[e.g.,][]{Cacciapouti22} and radial velocity searches \citep[e.g.,][]{Butler17}. Roman is expected to expand the population of exoplanets detected with microlensing considerably \citep{Penny19, Johnson20}, and thousands of planets are predicted to be detectable using multi-epoch astrometry in \Gaia{} DR4 \citep{Perryman14}. While the photometric, spectroscopic, and astrometric variability signals are orders of magnitude larger for systems with black holes than for exoplanets, stars with black hole companions are expected to be much rarer.

One technique used to search for exoplanets that has not been applied to black hole searches is direct imaging \citep[e.g.,][]{Marois08}. Here, we report results from a pilot survey combining RV selection with direct imaging using SHARK-VIS \citep{Pedichini22, Pedichini24} on the Large Binocular Telescope \citep[LBT,][]{Hill10}. Methods such as speckle imaging \citep[e.g.,][]{Davidson24}, interferometry \citep[e.g.,][]{Rizzuto13, Klement22, Deshmukh24}, and high-contrast imaging \citep[e.g.,][]{Shatsky02, Pauwels23} have long been used to identify and characterize binary systems and to search for wide companions to close binaries \citep[e.g.,][]{Isobe92, Tokovinin06}. These systems are important for understanding binary population statistics, which informs binary formation theories \citep[e.g.,][]{Moe17}. 

Direct imaging can be adapted to search for stars with black hole companions. If we resolve a system with large amplitude RV variability indicating a massive companion, but find no luminous star at the expected separation, the companion must be a compact object. Repeated observations can then measure the orbit of the photometric primary to determine the inclination. If a luminous star is identified, the component masses can be measured with two direct-imaging measurements at different orbital phases. Many of the binaries most suited for this observing strategy are bright ($V\lesssim 10$~mag), nearby stars. Even for systems where the companion is determined to be a star, rather than a compact object, direct imaging can provide the first characterization of the stellar companion, contributing to long-term efforts to understand stellar multiplicity. Section \S\ref{sec:target_selection} describes the search strategy for our program, and Sections \S\ref{sec:sharkvis_observations} and \S\ref{sec:lili} describe the properties of the instruments and the setup for our observations. We fit RV orbits using archival measurements in Section \S\ref{sec:rv_orbits} and show how the combination of high-contrast imaging and RVs can be used to directly measure the binary component masses. We characterize the binaries using broadband photometry in Section \S\ref{sec:sed_fits}, and we report our first results on four systems in Section \S\ref{sec:individual_targets}, and discuss how this search strategy could be applied to more targets in Section \S\ref{sec:discussion}.

\begin{figure}
    \centering
    \includegraphics[width=\linewidth]{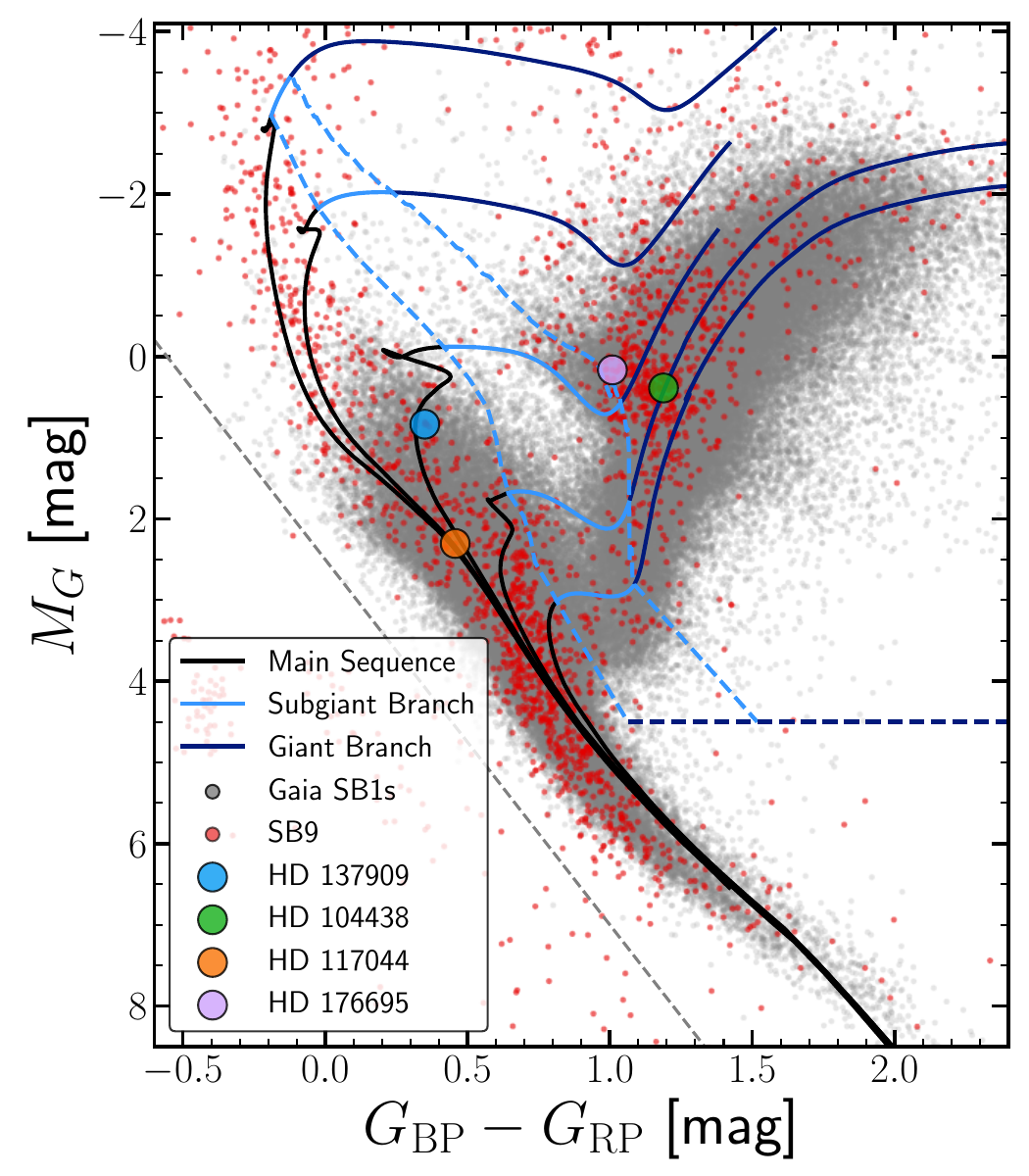}
    \caption{Extinction-corrected \Gaia{} CMD for targets in the SB9 catalog (red) and \Gaia{} SB1 catalog (gray). The black and blue lines show the CMD divisions into equal-mass main sequence, subgiant, and giant binaries following \citet{Rowan22} based on MIST isochrones and evolutionary tracks \citep{Choi16, Dotter16}. The dashed gray line shows the cut used to subdwarf and white dwarf binaries below the main sequence.}
    \label{fig:cmd}
\end{figure}

\newpage
\section{Target Search} \label{sec:target_selection}

\begin{figure}
    \centering
    \includegraphics[width=\linewidth]{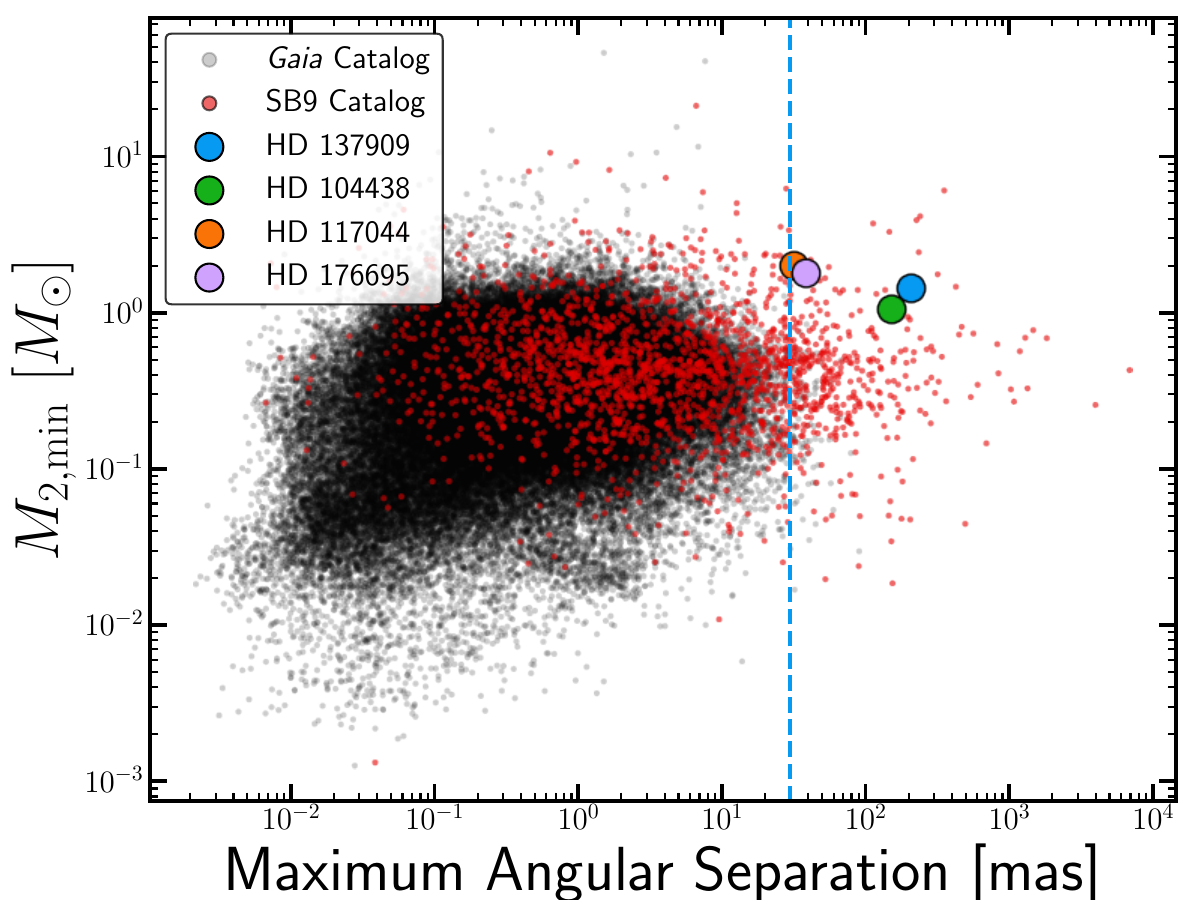}
    \caption{Minimum companion mass and maximum angular separation of the binaries in our search sample assuming an edge-on inclination. The four targets we observed with SHARK-VIS are labeled. The vertical blue line shows the angular separation where we expect to be able to confidently detect a companion with SHARK-VIS (30~mas).}
    \label{fig:selection_figure}
\end{figure}

\begin{table*}[]
    \centering
    \caption{Parameters of the SHARK-VIS observations. The DIT column reports the single-frame detector integration time.}
    \setlength{\tabcolsep}{4pt}
    \renewcommand{\arraystretch}{1.25}
    \begin{center}
        \begin{tabular}{c c c c c c c c c c}
        \toprule
        Target & Obs. Date & $R$ & Filter/$\lambda_{\rm{eff}}$ & Bandwidth & Plate scale & DIT & Total int. time & \% Selected frames & FoV Rotation\\
        & & (mag) & & (nm) & (mas/pix) & (ms) & (s) & & (deg) \\
        \midrule
        HD 137909 & 2024-05-17 & 3.5 & 647nm & 10 & 6.43 & 1.2 & 600 & 1.5 & 0.8\\
        HD 104438 & 2024-05-18 & 5.0 & H$\alpha/656.3$ & 5.0 & 4.32 & 10.0 & 300 & 100 & 8.0 \\
        HD 117044 & 2024-05-18 & 8.0 & $R/620$ & 135 & 4.32 & 2.3 & 630 & 3.0 & 11.1 \\
        -- & -- & -- & $V/500$ & 110 & 4.32 & 2.3 & 300 & 7.0 & 5.3 \\
        HD 176695 & 2024-05-17 & 7.4 & $R/620$ & 135 & 6.43 & 3.0 & 300 & 3.0 & 8.0 \\
        -- & -- & -- & $V/500$ & 110 & 6.43 & 3.0 & 300 & 3.0 & 8.0 \\
        \bottomrule
        \end{tabular}
    \end{center}
    \label{tab:sharkvis_setup}
\end{table*}

\begin{table*}[]
    \centering
    \caption{Properties of the detected companions in the SHARK-VIS images. Details on the observation setup are reported in Table \ref{tab:sharkvis_setup}. For HD~137909 and HD~104438 the flux ratios are measured in narrowband filters. For HD~117044 and HD~176695 we use $R$- and $V$-band filters.}
    \setlength{\tabcolsep}{4pt}
    \renewcommand{\arraystretch}{1.25}
    \begin{center}
        \begin{tabular}{l c c c c }
        \toprule
        & HD 137909 & HD 104438 & HD 117044 & HD 176695 \\
        \midrule
        Flux Ratio & $0.182\pm0.001$ & $(1.90\pm0.02)\times10^{-4}$ & $0.226\pm0.002$ & $(9.14\pm0.02)\times10^{-2}$\\
        $V$-band Flux Ratio & -- & -- & $0.2119\pm0.0005$ & $(10.0\pm0.6)\times10^{-2}$\\
        Separation & $130.56\pm0.05$ & $107.3\pm0.4$ & $30.42\pm0.50$ & $42.54\pm0.02$ \\
        Position Angle & $-92.34\pm0.01$ & $30.4\pm0.2$ & $161.86\pm0.07$ & $-19.84\pm0.03$ \\
        Method Used & Direct Imaging & PCA-ADI & Direct Imaging & Direct Imaging \\
        \bottomrule
        \end{tabular}
    \end{center}
    \label{tab:sharkvis_results}
\end{table*}

We selected candidates using two catalogs of spectroscopic binaries. The first is the SB9 catalog \citep{Pourbaix04}, which has spectroscopic orbits for more than 2300 systems observed as early as 1901 \citep{Crawford01}. We select systems with no reported secondary velocity semi-amplitude as the SB1 sample. Many of the binaries in the SB9 catalog are naked-eye stars, and the median $V$-band magnitude of the catalog is $8.3$~mag.

The second is the SB1 catalog in \Gaia{} DR3, which includes $>$180,000 systems \citep{GaiaHiddenTreasure, Gosset24}. The \Gaia{} RVS spectrometer operates in the near-IR (845--872~nm) around the CaII triplet and has $R\approx 11{,}500$ \citep{Cropper18}. The SB1 catalog has a median $G$~band magnitude of $11.9$~mag. Only the RV orbit model is included in \Gaia{} DR3 (the epoch RV measurements are expected to be released in DR4). Comparisons between the \Gaia{} RV orbit model and archival spectroscopy \citep{Bashi22} and photometry \citep{Rowan23} have found that roughly half of the solutions are unreliable. Here, we start with the entire \Gaia{} SB1 catalog and vet individual candidates on a case-by-case basis. 

Figure \ref{fig:cmd} shows both catalogs on a \Gaia{} color-magnitude diagram (CMD). We compute extinctions using the {\tt mwdust} Combined19 three-dimensional dust map \citep[][based on \citealt{Drimmel03, Marshall06, Green19}]{Bovy16}. For targets in the SB9 catalogs without \Gaia{} photometry, we estimate their \Gaia{} apparent magnitudes using the $V$ and $I$ or 2MASS $H$ and $K_s$ magnitudes\footnote{\url{https://gea.esac.esa.int/archive/documentation/GDR2/Data_processing/chap_cu5pho/sec_cu5pho_calibr/ssec_cu5pho_PhotTransf.html}}. We remove targets below the main sequence where one or both of the components are likely white dwarfs or subdwarfs. We also only include systems where the parallax $\varpi$ to parallax uncertainty $\sigma_\varpi$ ratio is $\varpi/\sigma_\varpi > 20$, since accurate distances are required to estimate the projected angular separation. 

We compute the binary mass function using the SB9 or Gaia RV orbit solution as
\begin{equation} \label{eqn:massfunction}
    f(M) = \frac{P K^3}{2\pi G}\left(1-e^2\right)^{3/2} = \frac{M_2^3 \sin^3i}{(M_1+M_2)^2},
\end{equation}
where $P$ is the orbital period, $K$ is the velocity semiamplitude, $e$ is the orbital eccentricity, $i$ is the orbital inclination, and $M_1$ and $M_2$ are the masses. The binary mass function is an absolute lower limit on the mass of the unseen companion, $M_2$, which corresponds to the limit with a zero-mass photometric primary and an edge-on inclination. We can place more meaningful minimum companion mass limits by estimating the mass of the photometric primary.

We use the StarHorse2 (SH2) catalog \citep{Anders22} to estimate the mass of the photometric primary. SH2 combines broadband photometry and \Gaia{} astrometry to estimate stellar parameters, such as effective temperature, metallicity and mass. The SH2 model assumes that the targets are single stars, but since these systems are observed as SB1s, we assume that the flux ratio is low enough that SH2 can provide a reasonable mass estimate. Only $\sim 63$\% of the stars in the SB9 catalog have SH2 mass estimates, as compared to $96$\% of the \Gaia{} SB1 catalog. We predict the masses of stars without SH2 mass measurements using their CMD position. We split the sample of stars with SH2 masses into 80\% for the training set and 20\% for the validation set. We stratify over the evolutionary state so that the fraction of main sequence, subgiant, and giant binaries in the training and validation sets is the same as in the total dataset. We fit a $k$-nearest neighbors model (KNN) using the extinction-corrected \Gaia{} color and absolute magnitude. As expected, this model is most effective on the main-sequence where stars of different mass are more easily separated in color and absolute magnitude. In the validation set, the root mean squared errors are $0.11\ M_\odot$, $0.26\ M_\odot$, and $0.30\ M_\odot$ for main sequence, subgiant, and giant stars, respectively. The binaries in the validation set with large mass errors are found at higher distances and extinctions compared to the full sample and are more concentrated in the Galactic plane. 

Next, we determine the minimum companion mass, $M_{2,\rm{min}}$ assuming an edge-on inclination and the SH2 mass of the photometric primary, $M_1$ with Equation \ref{eqn:massfunction}. We determine the maximum angular separation by combining $M_1$, $M_{2,\rm{min}}$, $P$, and the \Gaia{} distance estimate. Figure \ref{fig:selection_figure} shows the companion mass and the maximum angular separation of all of the targets. Targets with large angular separations and higher minimum companion masses are the most promising for searching for non-interacting black holes with SHARK-VIS. 

We selected four targets that are observable in the Northern hemisphere, bright ($R \lesssim 8$~mag), with $M_{2,\rm{min}}>1\ M_\odot$ and maximum angular separations of $>30$~mas for this pilot study. They are labeled in Figure \ref{fig:selection_figure} and Figure \ref{fig:cmd} shows their position on the \Gaia{} color-magnitude diagram. 

\begin{table*}[]
    \centering
    \caption{RV orbit fits to the four targets. $T_0$ is the periastron time. We fit for RV offsets when multiple datasets are used. For HD~137909, $\delta_{\rm{RV,1}}$, $\delta_{\rm{RV,2}}$, and $\delta_{\rm{RV,3}}$ correspond to the Lick, CASPEC, and DDO observations, respectively. For HD~104438, $\delta_{\rm{RV,1}}$, $\delta_{\rm{RV,2}}$, and $\delta_{\rm{RV,3}}$ correspond to CORAVEL, the Cambridge Spectrometer, and the Mount Wilson 60in observations, respectively. Finally, for HD~176695, $\delta_{\rm{RV,1}}$ corresponds to the DDO observations. The orbital solution for HD~137909 in the SB9 catalog only includes the RVs from \citet{Neubauer44}. Our updated RV orbit model, which includes more recent measurements, does not agree with the SB9 solution at the $1\sigma$ level (see Section \S\ref{sec:HD137909}).}
    \setlength{\tabcolsep}{12pt}
    \renewcommand{\arraystretch}{2}
    \begin{center}
        \begin{tabular}{lrrrr}
\toprule
{} &                  HD 137909 &               HD 104438 &            HD 117044 &                  HD 176695 \\
\midrule
Period [d]                  &           $3851^{+3}_{-3}$ &   $13000^{+100}_{-100}$ &   $1890^{+10}_{-10}$ &         $7710^{+40}_{-50}$ \\
$K$ [km/s]                  &     $9.34^{+0.06}_{-0.06}$ &     $6.7^{+0.6}_{-0.6}$ &     $30^{+40}_{-10}$ &       $11.6^{+0.2}_{-0.2}$ \\
Ecc                         &  $0.529^{+0.005}_{-0.005}$ &  $0.58^{+0.04}_{-0.05}$ &  $0.7^{+0.1}_{-0.1}$ &  $0.662^{+0.007}_{-0.007}$ \\
$f(M)\ [M_\odot]$           &  $0.199^{+0.004}_{-0.004}$ &  $0.22^{+0.07}_{-0.05}$ &    $1.3^{+7}_{-1.0}$ &     $0.52^{+0.02}_{-0.02}$ \\
$\gamma$ [km/s]             &   $-21.47^{+0.05}_{-0.05}$ &    $24.2^{+0.3}_{-0.3}$ &      $-12^{+2}_{-1}$ &    $-1.81^{+0.09}_{-0.09}$ \\
$\omega\ [^{\circ}]$        &      $179.5^{+0.4}_{-0.6}$ &       $-80^{+10}_{-10}$ &        $5^{+7}_{-3}$ &             $50^{+1}_{-1}$ \\
$T_0-2.42\times10^6$ [d]    &         $9000^{+10}_{-10}$ &    $1400^{+300}_{-200}$ &  $28560^{+30}_{-20}$ &       $13200^{+100}_{-80}$ \\
$\log s$ [km/s]             &             $-6^{+3}_{-3}$ &          $-6^{+3}_{-3}$ &       $-5^{+3}_{-3}$ &             $-6^{+3}_{-3}$ \\
$\delta_{\rm{RV},1}$ [km/s] &        $1.8^{+0.1}_{-0.1}$ &     $1.4^{+0.6}_{-0.7}$ &                    - &       $-0.3^{+0.9}_{-0.9}$ \\
$\delta_{\rm{RV},2}$ [km/s] &       $-0.2^{+0.1}_{-0.2}$ &     $1.1^{+0.7}_{-0.7}$ &                    - &                          - \\
$\delta_{\rm{RV},3}$ [km/s] &       $-1.1^{+0.1}_{-0.1}$ &           $0^{+1}_{-1}$ &                    - &                          - \\
\bottomrule
\end{tabular}

    \end{center}
    \label{tab:rv_table}
\end{table*}

\section{SHARK-VIS Observations} \label{sec:sharkvis_observations}

SHARK-VIS \citep{Pedichini22, Pedichini24} is a 400--1000~nm high-contrast optical imager using the right side SOUL Extreme Adaptive Optics (ExAO) system of the LBT \citep{Pinna16}. With an angular resolution of $\sim 15$~mas at 550nm \citep{Mattioli19}, it is the ideal instrument to search for binary companions to nearby SB1s. SHARK-VIS adopts a fast imaging approach with short integrations of a few milliseconds so that atmospheric turbulence distortions are roughly constant in each exposure \citep{Stangalini16}. Faint companions can then be detected with post-processing by recentering and stacking the frames to avoid smearing the PSF. The frame sequence is cleaned of detector signatures by a standard dark subtraction followed by frame-by frame removal of the dynamical column-amplifier noise measured using the masked pixels surrounding the field of view. The frames are then registered to the Gaussian fit peak of the PSF cores, numerically de-rotated to the sky orientation, and finally averaged, usually after selecting the best frames based on the peak amplitude and sharpness.

We observed the four targets during the nights of 2024-05-17 and 2024-05-18. Table \ref{tab:sharkvis_results} reports the details of the SHARK-VIS configuration for each target. For these observations, we used the ``LMIRCam'' entrance dichroic. All of the light below 700nm and half of the light between 700nm and 1000nm is sent to the SHARK-VIS instrument and the rest is sent to the AO wavefront sensor (WFS). The beam passes through the atmospheric dispersion compensator (ADC) and finally an optical filter. We used narrow-band filters for HD~137909 and HD~104438, and $R$- and $V$-band filters for HD~117044 and HD~176695. Since the WFS does not measure optical abberations along the optical path between the dichroic and the SHARK-VIS detector, a previously calibrated non-common path aberration correction must be inserted into the AO control loop. Unfortunately, this correction was not perfect for the observations presented here due to an issue with the AO setup, producing trefoil abberations in the image PSFs. This does not affect the first two targets (HD~137909, Section \S\ref{sec:HD137909} and HD~104438, Section \S\ref{sec:HD104438}), where the separation of the components is large ($\gtrsim 100$~mas). In the end, the trefoil abberation also did not hinder our analysis at all of the companions to HD~117044 (Section \S\ref{sec:HD117044}) or HD~176695 (Section \S\ref{sec:HD176695}). 

In all four systems we find evidence of a luminous companion. For three of our targets (HD~137909, HD~117044, and HD~176695), the companion is clearly visible in the stacked image even before selecting best frames. For the fourth target, HD~104438, angular differential imaging based on principal component analysis \citep[PCA-ADI,][]{Amara12} is required to detect the companion. We discuss the analysis of the SHARK-VIS images for each target individually in Section \S\ref{sec:individual_targets}. Only two of the targets of consistent with a simple stellar binary (HD~137909, Section \ref{sec:HD137909} and HD~117044, Section \S\ref{sec:HD117044}), and the other two are likely hierarchical triple systems (HD~104438, Section \S\ref{sec:HD104438} and HD~176695, Section \S\ref{sec:HD176695})

\section{Little iLocater (Lili)} \label{sec:lili}

During our SHARK-VIS observations, the iLocater pathfinder instrument, Little iLocater (Lili), was deployed on the left side of the LBT. Lili is a precursor to the full iLocater spectrograph that is currently under construction. The primary science objective of iLocater is to characterize exoplanets using extreme precision radial velocities \citep{Crepp16, Crass22}. iLocater will be diffraction limited and take advantage of the LBT AO system to get spatially-resolved spectroscopy, enabling characterization of close binary systems and exoplanet hosts. Both Lili and iLocater utilize the iLocater fiber injection system (acquisition camera) which was installed at the LBT in 2019 allowing for on-sky fiber coupling operations. Lili is a volume phase holographic grating spectrograph designed to validate the on-sky fiber-injection system for iLocater. Lili operates in the near-IR (0.97--1.31~$\mu$m) and has a resolving power of $R\approx1500$ recorded at speeds of up to 400~Hz. We observed three of our targets (HD~137909, HD~104438, and HD~176695) with Lili. For HD~137909, the companion was clearly detected in the iLocater acquisition camera, so we obtained spectra of both components. For HD~104438 and HD~176695 we only obtained spectra of the primary star.

The Lili data was reduced using the standard methodology described in Harris et al. (2024, submitted). In short, after standard image corrections, we optimally extracted the spectra using a halogen lamp flat field to define the spectral profile. We then wavelength-calibrated the spectra using an Ar lamp, and continuum normalized it using a spline fit. We also used the flat field lamp to remove a component of high frequency striping noise inherent to the InGaAs detector. For each target, we compare the observed Lili spectra to BT-Settl model spectra \citep{Allard11} matching the stellar parameters of each system. The model spectra are continuum normalized with a spline fit and convolved with a Gaussian function approximating the Lili line spread function to better match the data. We discuss the Lili spectra for each target in Section \S\ref{sec:individual_targets}. 

\section{RV Orbits} \label{sec:rv_orbits}

For each target, we fit the archival RVs using a Keplerian orbit model of the form
\begin{equation} \label{eqn:orbit}
        \text{RV}(t) = \gamma + K \left[(\omega+f)+e\cos\omega\right]
\end{equation}
\noindent where $\gamma$ is the center-of-mass velocity, $K$ is the radial velocity semiamplitudes, $f$ is the true anomaly, and $\omega$ is the argument of periastron. The true anomaly, $f$, is defined based on the eccentric anomaly, $E$, and the eccentricity, $e$ by
\begin{equation}
    \cos f = \frac{\cos E - e}{1-e \cos E},
\end{equation}
and the eccentric anomaly is
\begin{equation}
    E - e\sin E = \frac{2\pi(t-T_0)}{P}
\end{equation}
where $P$ is the period and $T_0$ is the time of periastron. We use {\tt emcee} \citep{ForemanMackey13} to sample over the orbital parameters with Markov Chain Monte Carlo (MCMC) methods. For targets with RVs from multiple instruments, we include an RV offset, $\delta_{\rm{RV}}$ to fit for differences in RV zero-points. We also fit for a stellar jitter term, $s$, which encompasses additional variation from stellar activity and the effects of underestimated RV uncertainties \citep{PriceWhelan17}. We run the MCMC chains for 50,000 iterations with 2$n$ walkers, where $n$ is the number of fit parameters. We discard the first 10,000 iterations as burn-in. Table \ref{tab:rv_table} reports the the new estimate of the mass function $f(M)$ and median values for all parameters with $1\sigma$ uncertainties from the MCMC posteriors. We discuss the RV orbit for each target in Section \S\ref{sec:individual_targets} and compare it to the original solution reported in the SB9 catalog. 

With a single SHARK-VIS measurement of the angular separation and an estimate of the photometric primary mass, we can measure the orbital inclination and the companion mass. A Keplerian orbit is defined by seven orbital elements, $P$, $T_0$, $e$, $\omega$, $i$, $a$ (semi-major axis), and $\Omega$ (longitude of ascending node). This first four, $P$, $T_0$, $e$, $\omega$ are measured directly from the RV orbit. The semi-major axis, $a$ is related to the masses $M_1$ and $M_2$ and the orbital period through Kepler's third law. If $M_1$ and $i$ are known, $M_2$ is determined by combining $K$, $P$, and $e$ (Equation \ref{eqn:massfunction}). The sky projected angular separation is 

\begin{equation} \label{eqn:projected_separation}
    \Delta\theta= \frac{a}{d}\frac{1-e^2}{1+e \cos f}\sqrt{1-\sin^2 i \sin^2(f+\omega)}
\end{equation}
\noindent where $d$ is the distance \citep{Tremaine23}. The sky-projected angular separation does not depend on $\Omega$, although $\Omega$ can be determined from the position angle of the separation. By combining the RV orbit and one measurement of the angular separation, we can determine $i$ and $M_2$ for an assumed value of $M_1$. With two measurements of the angular separation at different orbital phases, we can also measure $M_1$. We show an example of combining multiple angular separation measurements below in Section \S\ref{sec:HD137909}.

\begin{table}[]
    \centering
    \caption{Parameters of the spectral energy distribution for three of the four targets. For HD~137909, we adopt the values from \citet{Bruntt10}. We fit a single-star model to HD 104438 since we detect a wide tertiary rather than the inner companion with SHARK-VIS (Section \S\ref{sec:HD104438}). We find that HD~176695 is most likely a triple system, and discuss possible SED models for this system in Section \S\ref{sec:HD176695}.} 
    \setlength{\tabcolsep}{8pt}
    \renewcommand{\arraystretch}{2}
    \begin{center}
        \begin{tabular}{lrrrr}
\toprule
{} &                             HD 137909 &                              HD 104438 &                             HD 117044 \\
\midrule
$T_{\rm{eff},1}$ [K] & $7908\pm180$ & $4750^{+2}_{-1}$ & $7010^{+20}_{-10}$ \\
$T_{\rm{eff},2}$ [K] & $6750\pm230$ & - & $6750^{+50}_{-60}$ & \\
$\log g_1$           & $3.92\pm0.04$ & $2.54^{+0.02}_{-0.05}$ & $4.07^{+0.1}_{-0.07}$ \\
$\log g_2$           & $4.20\pm0.05$ & - & $4.8^{+0.2}_{-0.3}$\\
$L_1\ [L_\odot]$     & $25.3\pm2.9$ & $61^{+2}_{-2}$ & $7.71^{+0.03}_{-0.02}$ \\
$L_2\ [L_\odot]$     & $4.5\pm0.5$ & - & $1.705^{+0.009}_{-0.02}$ \\
$R_1\ [R_\odot]$     & $2.63 \pm 0.09$ & $11.5^{+0.2}_{-0.2}$ & $1.880^{+0.006}_{-0.007}$\\
$R_2\ [R_\odot]$     & $1.56 \pm 0.07$ & - & $0.96^{+0.02}_{-0.01}$\\
Distance [pc]        & $34.97$ & $96^{+2}_{-2} $& 143.10\\
$A_V$ [mag]          & 0.0 & $0.046^{+0.006}_{-0.005}$ & 0.0\\
Metallicity          & 0.28 & $-0.119^{+0.01}_{-0.006}$ & 0.0\\
\bottomrule
\end{tabular}
    \end{center}
    \label{tab:sed_table}
\end{table}

\section{Spectral Energy Distributions}  \label{sec:sed_fits}

We jointly fit the broadband photometry and flux ratio from the SHARK-VIS observations to characterize the binaries. We retrieve Galaxy Evolution Explorer \citep[\GALEX{},][]{Bianchi11}, \Gaia{} \citep{GaiaDR3}, Two Micron All Sky Survey \citep[2MASS,][]{Cutri03}, and Wide-Field Infrared Survey Explorer \citep[\WISE{},][]{Cutri12} photometry for our targets, when available. We also use \Gaia{} synthetic photometry \citep[GSPC,][]{Gaia23_gspc}, which is determined from the low-resolution \Gaia{} XP spectra. Since our targets are bright, some of the photometry is saturated. We discuss the quality of the individual photometry measurements for each target in Section \S\ref{sec:individual_targets}. 

We fit the spectral energy distribution (SED) defined by the available photometry using the \citet{Castelli03} model atmospheres included in {\tt pystellibs}\footnote{\url{https://github.com/mfouesneau/pystellibs}}. We use {\tt emcee} \citep{ForemanMackey13} to sample over the stellar parameters of both stars. We assume a fixed distance from \citet{BailerJones21} and extinction from {\tt mwdust}. We also keep the metallicity fixed and use literature metallicity measurements when available. We fit for two stellar components simultaneously, sampling over the effective temperatures, $T_{\rm{eff},1}$ and $T_{\rm{eff},2}$, surface gravities, $\log g_1$ and $\log g_2$, total luminosity, $L_T = L_1 + L_2$, and luminosity ratio $L_2 / L_1$. Finally, we include a prior on the flux ratio based on the SHARK-VIS observations in the filter used for each target. For HD~117044 and HD~176695, we use a prior on both $R$- and $V$-band flux ratios. We run the MCMC for 15,000 iterations with 12 walkers and discard the first 5,000 iterations as burn-in. Table \ref{tab:sed_table} reports the results of the two-star SED fits. We discuss the SED fits below. 

\section{Discussion of Individual Targets} \label{sec:individual_targets}

\subsection{HD 137909} \label{sec:HD137909}

\customfigureilocator{HD137909}{Panel (a): SHARK-VIS image of HD 137909 using the 647~nm narrowband filter. The companion is clearly visible at a separation of $\Delta\theta = 130.56$~mas with a flux ratio of $\alpha=0.182$. Panel (b): archival RVs from \citet{Neubauer44}, \citet{Kamper90}, \citet{North98}, and \citet{Mathys17}. 100 random samples of the RV orbit posteriors are shown in red. Panel (c): Constraints on the companion mass and orbital inclination from the RV orbit (black), SHARK-VIS (red), and VLT/NACO (blue) observations. Solid, dashed, and dotted lines show different values of $M_1$. By combining all three observations, we can uniquely determine $M_1=2.04\ M_\odot$, $i=65^{\circ}$, and $M_2=1.49\ M_\odot$, and this solution is marked as the cross. Panel (d): Multi-band photometry and two-star SED model from \citet{Bruntt10}. The blue square shows the expected flux of the secondary based on the flux of the photometric primary and the SHARK-VIS flux ratio. Panel (e): Lili spectra of the primary (blue) and secondary (red). The black lines show BT-Settl model atmospheres for both components. Both components have Paschen series lines and the companion has evidence for SI and CI absorption lines.}

HD 137909 ($\beta$ CrB) is a bright ($V=3.7$ mag) Ap main sequence star (Figure \ref{fig:cmd}). HD 137909 was first identified as a spectroscopic binary in 1907 and the first orbit was published in \citet{Cannon12} using 153 spectra taken at the Dominion Observatory between 1910 and 1912\footnote{\citet{Cannon12} found an orbital period of 490.8 days with a secondary period of 40.9 days}. \citet{Neubauer44} reports 341 RV measurements taken between 1931 and 1942 with the Mills spectrograph at the Lick Observatory \citep{Campbell1898}. \citet{Neubauer44} finds an orbital solution of $P=10.496$~years and $K=9.19$~km/s, and this is the orbital solution reported in the SB9 catalog \citep{Pourbaix04}.

HD 137909 has since been targeted by a variety of spectrographs. \citet{Kamper90} collected 121 RVs from the David Dunlap Observatory (DDO). \citet{North98} took 78 RV measurements of HD 137909 with the CORAVEL spectrograph on the 1m telescope at the Olservatoire de Haute-Provence. HD 137909 was later observed by \citet{Mathys17} as part of a program to measure magnetically split lines of Ap stars using CASPEC on the ESO 3.6m telescope, the AURELIE spectrograph on the 1.5m telescope at Observatoire de Haute-Provence (OHP), and the Kitt Peak National Observator (KPNO) Coud\'e spectrograph. We include all 67 measurements in our RV fit and use a single RV offset for the \citet{Mathys17} RVs, since they do not report which observations correspond to which instrument. There are also 93 RVs of this target from HARPS \citep{Trifonov20} from 2007, but they are not zero-point corrected and were all taken in the same night, so we do not include them in our orbit fit.

Figure \ref{fig:panel_HD137909}b shows the RVs and random samples from the MCMC posteriors and Table \ref{tab:rv_table} reports the RV orbit fit from the combined RV datasets. Our RV orbit fit prefers a slightly longer orbital period, $3851\pm3$~days ($10.55$~years), than the SB9 catalog orbit $P=3833.58$~days, which is only based on the \citet{Neubauer44} RVs. We also measure a higher orbital eccentricity, $0.529\pm0.005$ versus $0.406\pm0.025$. Our updated RV model leads to a lower binary mass function of $f(M) = 0.199\pm0.004\ M_\odot$ instead of the $f(M)=0.24\ M_\odot$ reported in \citet{Neubauer44}.

HD 137909 is not included in the StarHorse2 catalog, which uses \Gaia{} DR3 parallaxes and magnitudes, but it was included in the original StarHorse catalog \citep{Anders19} based on \Gaia{} DR2. The StarHorse mass is $1.5^{+0.3}_{-0.4}\ M_\odot$. The \Gaia{} DR3 mass estimate from the FLAME model pipeline is significantly higher with $M_1=2.14\pm0.05\ M_\odot$. 

The primary is a main sequence star, so the effective temperature should be a good proxy to estimate the stellar mass. However, there are a number of disparate effective temperature estimates in the literature. \citet{Gray03} measured $T_{\rm{eff}}=7624$~K using spectra from the 0.8m telescope at the Dark Sky Observatory. The ELODIE spectra library reports $T_{\rm{eff}}=8763$~K \citep{Prugniel07}. A reanalysis of the archival ELODIE spectra using a principal component analysis based on a set of spectroscopic standard stars found a lower temperature from the combined set of ELODIE spectra of $T_{\rm{eff}}=7635\pm82$~K \citep{MunozBermejo13}. \citet{Prungjel11} used the MILES medium-resolution spectrograph to find $T_{\rm{eff}}=8466\pm277$~K. The HARPS radial velocity database \citep{Perdelwitz24} includes a spectroscopic fit using SPECIES \citep{Soto18} with $T_{\rm{eff}}=7500\pm300$~K for HD 137909. A typical A9/F0 star ($M\sim1.7\ M_\odot$) has $T_{\rm{eff}}=7200$--$7400$~K \citep{Pecaut13}, which is cooler than any of the spectroscopic effective temperature measurements. 

Since the luminous companion was not detected in the spectra used to derive the RVs, we observed this target with SHARK-VIS to search for evidence of a luminous companion. However, we later found that \citet{Bruntt10} had previously discovered the companion using VLT/NACO adaptive optics imaging.

We observed HD 137909 with SHARK-VIS on 2024-05-17 with a plate scale of 6.43~mas/pix and a narrow-band 647~nm filter. We used 1.2~ms detector integration times for a total exposure of 600s. Figure \ref{fig:panel_HD137909}a shows the sky-aligned, stacked image. The companion is clearly visible without using ADI to subtract the photometric primary. In post-ExAO high-resolution imaging, the PSF is characterized by a very complex shape where the diffraction core is surrounded by a number of variable and quasi-static speckles \citep[e.g.,][]{Stangalini16}. The lack of a smooth background makes reliable photometric and astrometric measurements within the AO control radius ($\sim 250$ -- 300~mas, Figure \ref{fig:panel_HD137909}a) challenging. We characterize the companion by taking a Bayesian modeling approach based on numerical information field theory \citep[NIFTy,][]{Arras19}. We maximize the posterior distribution for the unknown PSF, $A(x,y)$, and companion location, $(dx,dy)$, and the flux ratio, $\alpha$, 
\begin{equation}
    I(x,y) = A(x,y) + \alpha A(x-dx, y-dy),
\end{equation}
\noindent where $I(x,y)$ is the stacked image after selecting the best 1.5\% of frames. We find a companion flux ratio of $0.182\pm0.001$ and a separation $130.56\pm0.05$~mas (Table \ref{tab:sharkvis_results}). The separation is less than the predicted maximum angular separation used to select this target as a candidate ($205$~mas). The detection of a stellar companion means a non-interacting companion can be ruled out. 

By combining our measurement of the SHARK-VIS angular separation with the earlier observation from VLT/NACO \citep{Bruntt10} with the RVs, we can break the degeneracy between the primary mass and the inclination and measure the companion mass. Figure \ref{fig:panel_HD137909}c shows the constraints on $M_2$ as a function of $i$ from the RV orbit, SHARK-VIS observation, and VLT/NACO observation. We find that $M_1=2.04\ M_\odot$ and $i=65^{\circ}$, and the resulting companion mass is $M_2=1.49\ M_\odot$.

Figure \ref{fig:panel_HD137909}d shows the SED of HD 137909. The 2MASS IR photometry is saturated, so we instead use $J$ and $K$ band magnitudes from \citet{Morel78}. The \WISE{} $W1$ photometry is also saturated. Figure \ref{fig:panel_HD137909}d shows the SED model from \citet{Bruntt10}. Since they use the Hipparcos parallax $\varpi = 28.60\pm0.69$, we use the same value here rather than the zeropoint-corrected \Gaia{} DR3 parallax $\varpi = 27.96\pm0.97$. The model under predicts the flux in the \Gaia{} synthetic photometry, but the apparent $G$-band magnitude is $3.6$~mag, and the XP spectra for stars $G \lesssim 5$~mag could be affected by saturation \citep{Gaia23_gspc}. The flux ratio of the model at 647~nm is $\alpha=0.19\pm0.03$, which is consistent with the measured SHARK-VIS flux ratio. The SED parameters are reported in Table \ref{tab:sed_table}.

Figure \ref{fig:panel_HD137909}e shows the Lili spectrum of each component. BT-Settl model atmospheres \citep{Allard11} are shown for comparison using the effective temperature, surface gravity, and metallicity reported in Table \ref{tab:sed_table}. Both components have Paschen lines, and there is evidence for SI and CI lines in the companion, but these features are absent in the spectrum of the primary. 

In summary, HD~137909 is a long period ($10.55$~years) binary with two main sequence stars. By combining our SHARK-VIS measurement with the VLT/NACO observation from \citet{Bruntt10}, we measure the component masses to be $2.04\ M_\odot$ and $1.49\ M_\odot$ (Figure \ref{fig:panel_HD137909}c).

\subsection{HD 104438} \label{sec:HD104438}

\customfigureilocator{HD104438}{Panel (a): SHARK-VIS image of HD 104438 taken with an H$\alpha$ filter after using PCA-ADI to remove the photometric primary. The white circle traces the AO control radius, and the bright features along this line are instrumental artifacts. A luminous companion is detected at a separation of 107~mas. Panel (b): archival RVs from \citet{Harper23}, \citet{Griffin08}, and APOGEE. The vertical line marks the time of the SHARK-VIS observation. Panel (c): constraints on the mass of the unseen companion mass and orbital inclination assuming $M_1=1.22\ M_\odot$. The black curves show the constraint from the binary mass function (Equation \ref{eqn:massfunction}) and the red curves show the constraint from the projected angular separation (Equation \ref{eqn:projected_separation}). The two curves intersect for a high companion mass, which is inconsistent with the observed flux ratio, so the observed wide companion is likely a tertiary companion. For comparison, the blue and purple curves show the same curves if we had detected a companion at a separation of 30, 40, or 45~mas. Panel (d): single-star SED fit to the optical and IR photometry of the red giant. The blue square shows the measured SHARK-VIS flux ratio. The red curve shows the SED corresponding to a M-dwarf, and the blue curves show WD model atmospheres for a range of cooling ages. Panel (e): Lili spectrum of the photometric primary (blue) and BT-Settl model atmosphere (black). In addition to the Paschen lines, there are some metal lines between 1000--1100~nm.}

\begin{figure*}
    \centering
    \includegraphics[width=\linewidth]{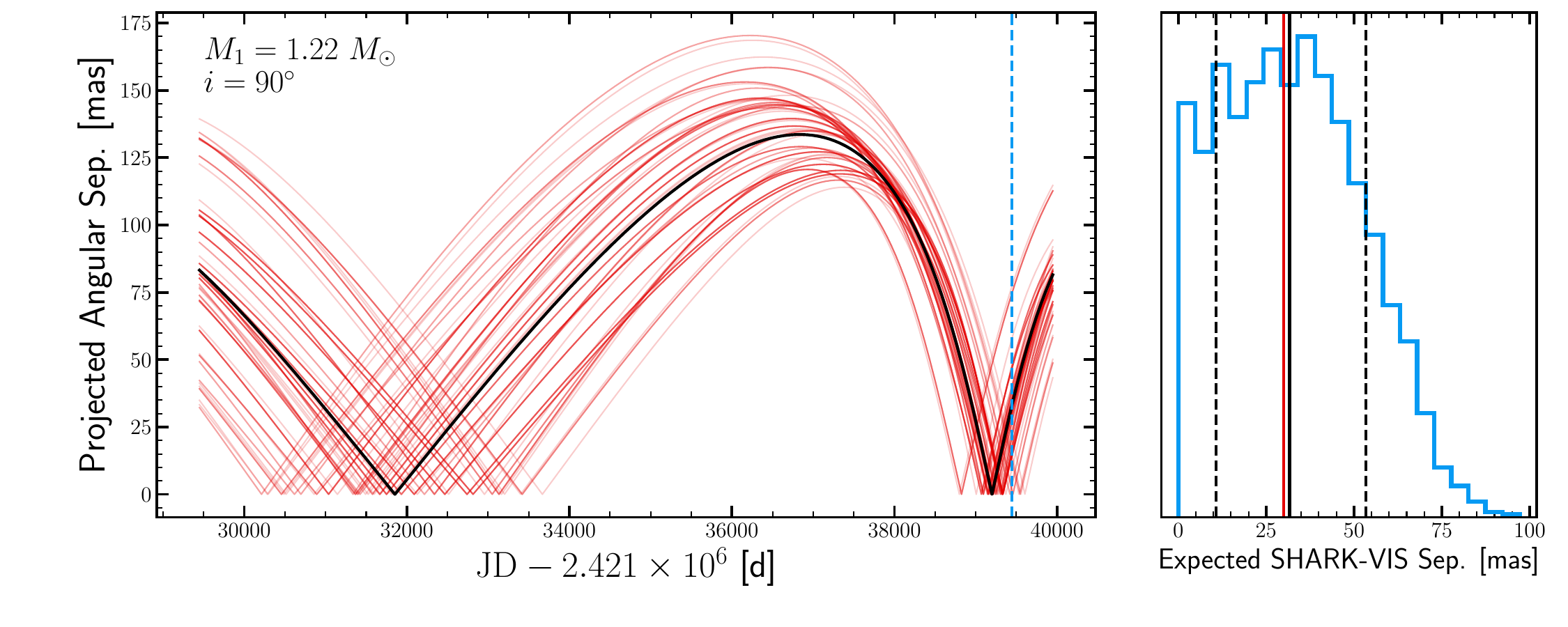}
    \caption{Left: projected angular separation as a function of time for HD 104438 assuming $M_1=1.22\ M_\odot$ and $i=90^\circ$. The black line corresponds to the RV solution using the median of the MCMC posteriors, and the red lines are random samples from the posteriors. The vertical blue line indicates the time of the SHARK-VIS observation. Right: distribution of predicted angular separations at the SHARK-VIS observation for $M_1=1.22\ M_\odot$ and $i=90^\circ$. The vertical black lines mark the median and 16th and 84th quantiles and the red line shows the nominal detection limit for our SHARK-VIS observation. Additional RVs are needed to improve the orbit ephemeris and constrain the predicted angular separation.}
    \label{fig:HD104438_angular_sep}
\end{figure*}

HD 104438 is an $R\approx 5.0$~mag red giant (Figure \ref{fig:cmd}). The \Gaia{} photogeometric distance is $95.5$~pc \citep{BailerJones21}. HD 104438 was first characterized as a possible RV variable in the Bright Star Catalog \citep{Hoffleit82} based on RV measurements from \citet{Harper23}\footnote{As \citet{Griffin08} points out, this classification was based on a typo in the original manuscript from \citet{Harper23}.}. The SB9 catalog includes 75 RVs, the majority of which are from the CORAVEL spectrograph at Observatoire de Haute Provence or the Cambridge CORAVEL spectrograph obtained between 1992 and 2008 by \citet{Griffin08}. Since \citet{Griffin08} does not report RV uncertainties, we assume $\sigma_{\rm{RV}}=1.0$~km/s for the CORAVEL observations, $\sigma_{\rm{RV}}=2.5$~km/s for observations taken with the Cambridge Spectrometer, and $\sigma_{\rm{RV}}=5.0$~km/s for observations taken with the Mount Wilson 60in \citep{Adams23, Abt70, Harper23}. As described above, we include a term for stellar jitter in our RV orbit fits, which can account for underestimated RV uncertainties. HD 104438 was also observed by APOGEE in 2013, extending the baseline of the RV observations. Figure \ref{fig:panel_HD104438}b shows the RV observations and the RV orbit model. The orbital period is $P=13,000\pm 100$~d ($35.6\pm 0.3$~yr) and the mass function is $f(M)=0.22^{+0.08}_{-0.05}\ M_\odot$. This is consistent with the solution reported in the SB9 catalog. 

HD 104438 is included in the StarHorse catalog \citep{Anders19} with a mass of $M_1 = 1.3\pm0.3\ M_\odot$. The APOGEE catalog reports an effective temperature of $T_{\rm{eff}}=4733\pm8$~K, a surface gravity of $\log g=2.58\pm0.02$, and a metallicity of $[\rm{M}/\rm{H}]=-0.08$. \citet{Ting18} used APOGEE spectra to estimate asteroseismic parameters $\Delta P$ and $\Delta\nu$ and predicted that HD~104438 is a red clump star. \citet{Martig16} also used the APOGEE spectrum of HD~104438 to predict its mass based on its C and N abundances and found $M=1.22\ M_\odot$. Figure \ref{fig:panel_HD104438}c shows the companion mass $M_2$ for HD 104438 as a function of the orbital inclination for three different values of $M_1$. The unseen companion would only be massive enough to be a black hole with $M_2 > 3\ M_\odot$ if the orbital inclination is $\lesssim 30^{\circ}$. In order for the companion to be a neutron star with $M_2>1.4\ M_\odot$, the inclination would have to be $<55^{\circ}$ for $M_1 = 1.2\ M_\odot$, which corresponds to a probability $P(i < 55^{\circ}) = 0.42$ assuming an isotropic distribution of inclination angles. Based on the RV orbit, primary mass estimate, and \Gaia{} distance, we predict that the maximum angular separation is $148$~mas (Figure \ref{fig:selection_figure}), making this an ideal target for SHARK-VIS. 

HD 104438 was observed with SHARK-VIS on 2024-05-08 in an H$\alpha$ filter with a 5nm bandwidth. We used a larger $1.5\times$ magnification lens to get a plate scale of $4.32$~mas/pix. The total exposure time was 300s with a detector integration time of 10ms and a field of view rotation of 8~deg, allowing ADI to be used to search for a faint companion. The combination of fast imaging with a narrow band filter and the focal extender was possible because the source is bright ($G=5.29$~mag).

After subtracting a dark image and dynamical bias correction, all 30,000 frames were co-registered and processed by a principle component analysis based ADI \citep[PCA-ADI][]{Amara12} implemented through the IncrementalPCA {\tt scikit-learn} code \citep{Pedregosa11}. This revealed a faint companion at a separation of $\sim 100$~mas (Figure \ref{fig:panel_HD104438}a). Forty principal components were chosen to maximize the SNR of the companion. We use the NFC technique \citep{Marois10}, the standard for ADI photometry, to accurately measure the contrast and position of the companion by injecting negative scaled copies of the primary until the the companion is canceled. We found a contrast ratio of $(1.90\pm0.02) \times 10^{-4}$ and a separation of $107.3\pm0.4$~mas (Table \ref{tab:sharkvis_results}). To confirm that the detection is from a stellar companion rather than an instrumental artifact, we inject fake companions at the same angular separation and different position angles with contrast ratios $2\times10^{-4}$. The injected sources have similar shapes and brightness to the detected companion. We also run PCA-ADI on the first and second half of the images independently and find the companion at the same position in both data segments. 

Figure \ref{fig:panel_HD104438}b shows the RV orbit and marks the epoch of the SHARK-VIS observation. HD 104438 was observed shortly after periastron, so the separation should be much less than the maximum separation predicted for an edge-on orbit (148~mas). Even if the orbit had $i\sim 30^\circ$, we would only expect the companion to be separated by $\sim 70$~mas at the epoch of the SHARK-VIS observation. Figure \ref{fig:panel_HD104438}c shows the constraints on the companion mass using the RV orbit and the SHARK-VIS projected separation of 107~mas. The two curves intersect at $\sim 17^\circ$, which would require $M_2\sim 10\ M_\odot$ assuming $M_1=1.22\ M_\odot$. This solution is unphysical because a $10\ M_\odot$ star would dominate the SED and be inconsistent with the observed flux ratio. One explanation for the system could be that the faint companion seen with SHARK-VIS is a high mass ratio binary of a black hole and a low luminosity M-dwarf or white dwarf. However, it is more likely that the objected detected in SHARK-VIS is not the source of the observed RV variability, but is instead a wide tertiary or a chance alignment. We discuss both of these options below. 

The source of the RV variability in HD~104438 could either be a stellar companion that is too close to the giant to be resolved or a compact object. Figure \ref{fig:HD104438_angular_sep} shows the expected orbital separation as a function of time for an edge-on orbit and $M_1=1.22\ M_\odot$. For an edge-on orbit, we predict an angular separation of $30\pm20$~mas. The large uncertainty is driven by the poor ephemeris (Figure \ref{fig:panel_HD104438}b). Since the nominal separation detection limit is 30~mas for our SHARK-VIS observations, we cannot rule out a luminous companion at smaller separations. Additional RVs are needed to improve the ephemeris and provide tighter constraints on the range of possible inclinations. The orbital separation will continue to increase (Figure \ref{fig:HD104438_angular_sep}), so additional SHARK-VIS observations could be used to continue searching for the companion that is the source of the long-period RV variability.

The low flux-ratio source separated by 107~mas seen in SHARK-VIS could be a chance alignment of a fainter star or a wide tertiary companion. First, we fit a single-star SED to the red giant primary to estimate the total luminosity and estimate the possible spectral type of the faint source. We sample over the effective temperature, surface gravity, and luminosity of the giant, and the distance, extinction, and metallicity. We use a Gaussian prior on the distance based on the parallax uncertainty. We also include priors on the $T_{\rm{eff}}$ and $\log g$ of the giant from the APOGEE spectrum. As with the two-star SED models described in Section \ref{sec:sed_fits}, we use {\tt emcee} and run the MCMC for 15,000 iterations with 12 walkers and discard the first 5,000 iterations as burn-in. The 2MASS photometry is saturated in the $J$, $H$, and $K_s$ bands and has a ``D'' quality flag. Similarly, the \WISE{} $W1$ and $W2$ photometry is saturated with SNR$<2$. The $W3$ and $W4$ bands have higher SNR, (70.5 and 54.6, respectively) and are tagged as ``A'' quality photometry. The \GALEX{} NUV is saturated at $13.7$~mag, but we include the $FUV$ point in our analysis. We also include the Tycho $B$ and $V$ band measurements \citep{Wright03} in our fit. Figure \ref{fig:panel_HD104438}d shows the UV, optical, an IR photometry and the SED fit and Table \ref{tab:sed_table} reports the parameters and $1\sigma$ uncertainties. The detected companion is not intrinsically faint with $V\approx 15$~mag.

\begin{figure}
    \centering
    \includegraphics[width=\linewidth]{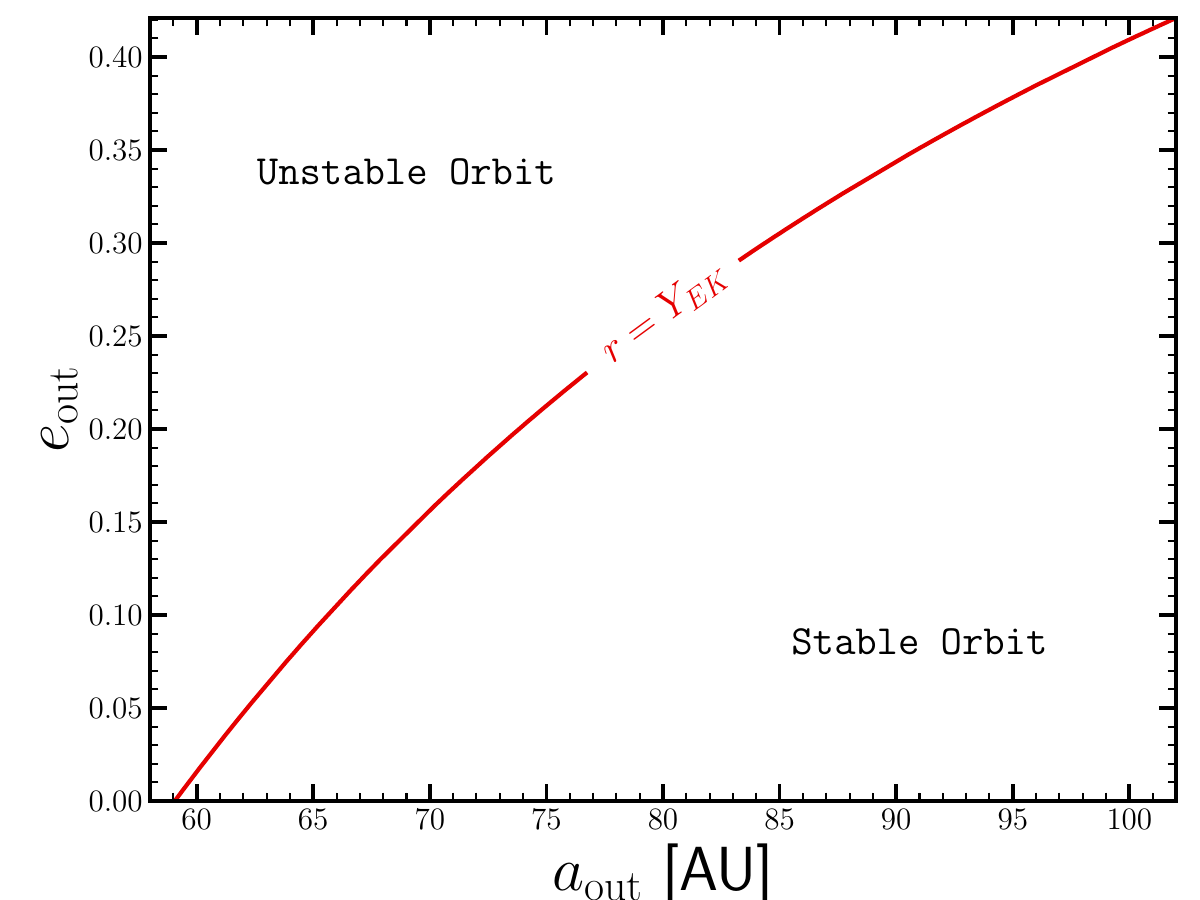}
    \caption{Orbital stability criterion for the HD~104438 triple system assuming $M_1 = 1.22\ M_\odot$ and $i=90^{\circ}$ for the inner binary. Orbits with $r/Y_{EK} > 1$ are stable, and the red line shows where $r/Y_{EK} = 1$. The faint companion seen in SHARK-VIS at 107~mas separation is consistent with a stable triple system if it is in a wide ($a_{\rm{out}}\gtrsim 60$~AU) orbit.}
    \label{fig:HD104438_orbital_stability}
\end{figure}

Figure \ref{fig:panel_HD104438}d also shows the SED of a $0.37\ M_\odot$ M-dwarf and WD cooling track models at the distance of HD 104438. We use \cite{Koester10} WD model atmospheres and cooling tracks from \citet{Bedard20}. Based on the low flux ratio of the SHARK-VIS detection, the wide companion is most likely a M-dwarf. We assess the orbital stability of such a triple using the criterion from \citet{Eggleton95} that the orbit is stable if $r>Y_{\rm{EK}}$, where
\begin{equation}
    Y_{EK} = 1 + \frac{3.7}{q_{\rm{out}}^{1/3}} - \frac{2.2}{1+q_{\rm{out}}^{1/3}} + \frac{1.4}{q_{\rm{in}}^{1/3}}\frac{q_{\rm{out}}^{1/3}-1}{1+q_{\rm{out}}^{1/3}},
\end{equation}
\noindent $q_{\rm{in}}=M_1 / M_2$, $q_{\rm{out}}=(M_1+M_2)/M_3$,
\begin{equation}
    r \equiv \frac{a_{\rm{out}}(1-e_{\rm{out}})}{a_{\rm{in}}(1+e_{\rm{in}})},
\end{equation}
and $a_{\rm{in}}$ and $a_{\rm{out}}$ are the semimajor axes and $e_{\rm{in}}$ and $e_{\rm{out}}$ are the orbital eccentricities. Figure \ref{fig:HD104438_orbital_stability} shows the ratio $r/Y_{EK}$ for a range of $e_{\rm{out}}$ and $a_{\rm{out}}$ assuming $M_1=1.22\ M_\odot$, $M_3=0.37\ M_\odot$, and that the inclination of the inner binary is edge-on. The detected tertiary is consistent with a stable hierarchical triple for wide orbits with $a_{\rm{out}} \gtrsim 60$~AU. We observe the system at a projected separation of $\sim 10$~AU, which suggests that the orbit is inclined. 

The probability of the detected source being a chance alignment of a background star is near zero. HD 104438 is at $l=169.7^{\circ}$, $b=76.2^{\circ}$ in Galactic coordinates, so the local stellar density is low. We calculate the probability of spurious association of a background star in two ways. First, we use \Gaia{} DR3 to determine the probability $P_s$ that a star of comparable brightness is found in a random $0.1$~mas radius patch corresponding to the angular separation of the M-dwarf. We draw 10,000 random circles with radius $0.1$~mas on the sky within $5^{\circ}$ of HD 104438 and count the number of stars in \Gaia{} DR3. We find no matches even without imposing an apparent magnitude cutoff, so we place a limit of $P_s < 1\times10^{-4}$. 

We also assess the probability of spurious association using the Besan\c{c}on model \citep{Czekaj14}, which is a stellar population synthesis model of the Milky Way that can produce an estimate of the star counts along the line of sight. The model has 131 stars with $V<15$~mag within the 1~deg patch towards HD 104438, so the probability that there is a random star in a given 0.1~mas region is $P_s \sim 1\times10^{-7}$. It is therefore unlikely that the detected star is a chance alignment, but additional SHARK-VIS observations are needed to confirm the tertiary companion. 

\begin{figure}
    \centering
    \includegraphics[width=\linewidth]{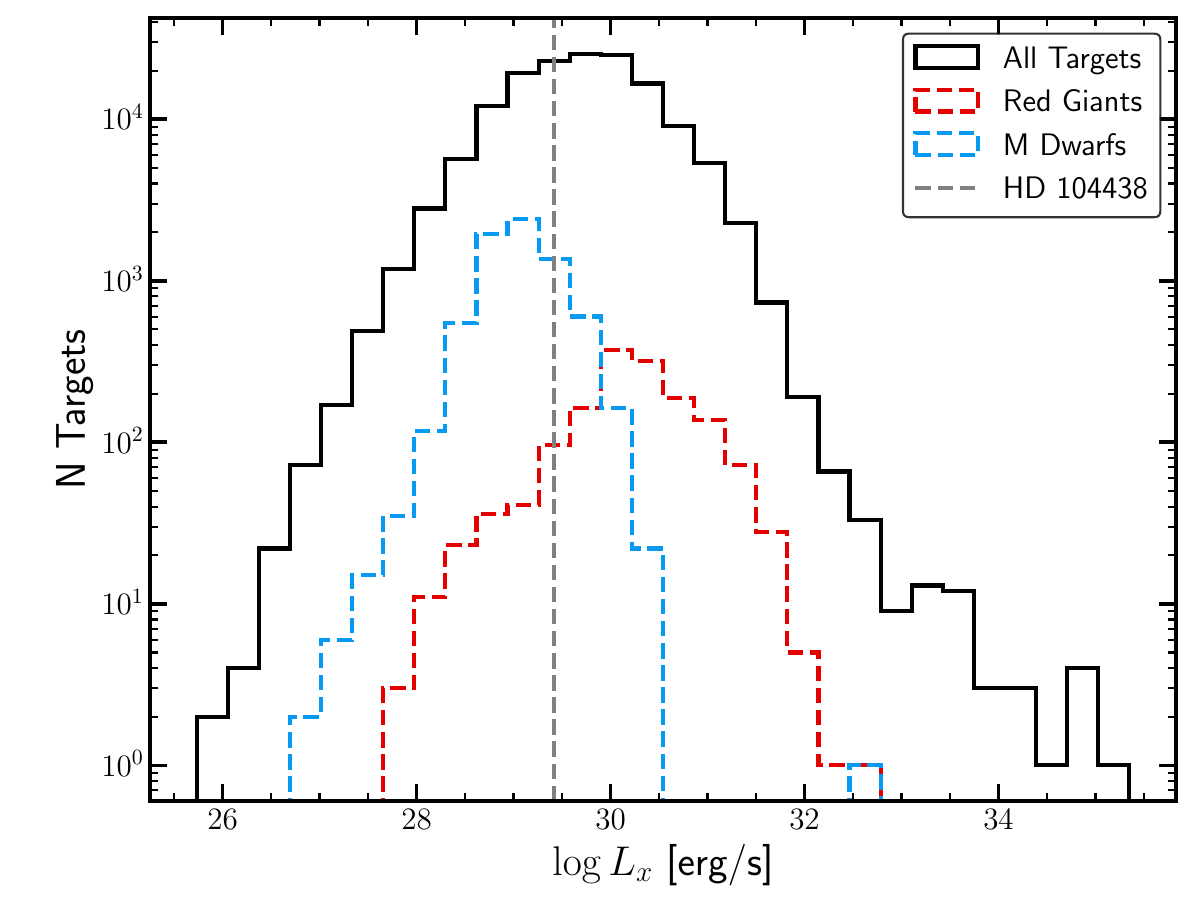}
    \caption{Distribution of X-ray luminosities for targets in the \textit{eROSITA} coronal line emitter catalog \citep{Freund24}. We select a subset of red clump targets near HD 104438 on the CMD within $|\Delta M_G|<0.1$ and $|\Delta(G_{\rm{BP}} - G_{\rm{RP}})| < 0.5$ (blue) and a comparison sample of M-dwarfs with $9.5 < M_G < 10.5$ and $2.6 < G_{\rm{BP}} - G_{\rm{RP}} < 3.1$ (red). The vertical line shows the ROSAT X-ray luminosity for HD 104438.}
    \label{fig:xray_hist}
\end{figure}

ROSAT identified HD 104438 as an X-ray source with an separation of $16\farcs6$ \citep{Freund22} from the optical source. The X-ray luminosity computed using the \Gaia{} distance is $L_x = 2.6\times10^{29}$~erg/s. Figure \ref{fig:xray_hist} shows the distribution of \textit{eROSITA} coronal line emitter X-ray luminosities \citep{Freund24} compared to $L_x$ for HD~104438. The X-ray emission could come from either the giant primary or the M-dwarf tertiary, but it is more consistent with the M dwarf population, and M dwarf X-ray emitters are more common than red giant X-ray sources.

\customfigure{HD117044}{Panel (a): SHARK-VIS $R$-band image of HD 117044. The companion is visible at a separation of 30~mas with a flux ratio of 0.226 (Table \ref{tab:sharkvis_results}). Panel (b): archival RVs from \citet{Halbwachs12}. As compared to the other three targets, the orbit is less well-constrained, especially near periastron, so the posteriors on $K$ and $f(M)$ are large (Table \ref{tab:rv_table}). Panel (c): constraints on the companion mass for different values of $f(M)$. Panel (d): two-star SED fit to the multi-band photometry with a prior on the $R$- and $V$-band flux ratios from SHARK-VIS.}

Finally, we obtained a spectrum of the giant primary using Lili. Since the flux ratio is low, the companion was not identified in the iLocater acquisition camera and no fiber was placed on the the tertiary. Figure \ref{fig:panel_HD104438}e shows the Lili spectrum of the photometric primary compared to a BT-Settl model spectra for a giant using the APOGEE effective temperature, surface gravity, and metallicity. The spectrum has a strong H~Pa~$\beta$ line and some possible metal lines between 1050--1100~nm.

We conclude that HD~104438 is most likely a hierarchical triple system. The inner binary contains the red giant photometric primary and the secondary is undetected in SHARK-VIS. Additional RVs are needed to refine the ephemeris and place limits on a lumionus companion.  The SHARK-VIS observations reveal a low flux ratio, $1.9\times10^{-4}$ (Table \ref{tab:sharkvis_results}, Figure \ref{fig:panel_HD104438}) star that must be the tertiary based on the phase of the inner binary RV orbit. This tertiary is most likely an M-dwarf and is unlikely to be the chance-alignment of a background star. Additional SHARK-VIS observations are necessary to detect the search for the RV component of the inner binary and confirm the tertiary. 

\subsection{HD 117044} \label{sec:HD117044}

HD 117044 is an A-type main sequence star \citep{Renson09} at a distance of 143~pc \citep{BailerJones21}. \citet{Halbwachs12} measured 45 RVs for this target with CORAVEL, and no other archival RVs are available. Figure \ref{fig:panel_HD117044}b shows the RV orbit of HD 117044. The RVs are consistent with a long-period eccentric orbit, but the velocity semiamplitude is uncertain because of the lack of observations near periastron. We fit the RV observations with a broad Gaussian prior on $K$ with a mean of 14.6~km/s, corresponding to half the range of the observed RV time series, and a standard deviation of 10 km/s. As expected, the RV posteriors are asymmetric, and RV orbits with high $K$ cannot be ruled out. Our RV orbit solution agrees with the solution reported in the SB9 catalog. It is possible that this system has a large $f(M) > 1\ M_\odot$, but additional RV observations are needed to confirm the orbit and better constrain $K$. Unfortunately, the next periastron passage is not expected until January 2028.  

The Starhorse2 catalog \citep{Anders22} reports $M_1 = 1.55^{+0.11}_{-0.06}\ M_\odot$. Figure \ref{fig:panel_HD117044}c shows the constraint on $M_2$ as a function of the orbital inclination for three different values of $f(M)$.  If $f(M) = 0.3\ M_\odot$, the companion could have $M_2 > 3.0\ M_\odot$ for inclinations $i < 37.8^{\circ}$. The secondary would be more massive than the primary unless the inclination is $ i < 66.7^{\circ}$. At higher mass functions, the companion mass could be very large, making this an ideal target for SHARK-VIS to search for a luminous companion. Based on the RV orbit and primary mass estimate, the maximum angular separation is $32$~mas.

We observed HD 117044 with SHARK-VIS on 2024-05-18 in the $R$- and $V$-band. The total exposure times were 630s and 300s in the $R$- and $V$-band, respectively, and both used a 2.3ms detector integration time. Since the companion is expected to be separated by $\sim30$~mas (Figure \ref{fig:selection_figure}), we used the $1.5\times$ magnification lens in order to maximize the plate scale (4.32~mas/pix). The companion is clearly visible in the stacked image made by averaging the 3\% of frames selected to have the largest Strehl ratio (Figure \ref{fig:panel_HD117044}a). The trefoil abberation produces three lobes around both stars, which complicates the flux ratio determination, so we applied the same Bayesian modeling approach used for HD~137909 (Section \ref{sec:HD137909}) to characterize the companion. The $R$-band flux ratio is $0.226\pm0.002$ and the separation is $30.42\pm0.50$~mas (Table \ref{tab:sharkvis_results}). The $V$-band flux ratio is $0.2119\pm0.0005$, and the uncertainties include both the optimization uncertainties and the uncertainties from the choice of box position. 

HD 117044 is the faintest of the four targets in $R$-band, and the broadband photometry is less affected by saturation in the near-IR. The 2MASS $J$, $H$, and $K_s$ magnitudes all have an ``A'' quality flag, as do the \WISE{} $W1$--$W4$ magnitudes. HD 117044 has also been observed by \GALEX{} and has NUV$=13.0$~mag, but since \GALEX{} photometry is saturated and non-linear for $\rm{NUV}<14.0$~mag, we exclude it from our SED fit. Figure \ref{fig:panel_HD117044}d shows the optical to IR photometry of HD 117044. We assume the metallicity is Solar since there is no previously reported spectroscopic metallicity measurement. We fit a two-star SED model with a prior on the $R$-band and $V$-band flux ratios measured by SHARK-VIS. Our SED fit (Table \ref{tab:sed_table}) prefers a solution with two main sequence stars. The primary has an effective temperature $T_{\rm{eff,1}}=7000$~K and radius $R_1=1.9\ R_\odot$ and the secondary has $T_{\rm{eff},2}=6800$~K and $R_2=0.96\ R_\odot$ suggesting that the primary is near the end of its main sequence lifetime.

In summary, we find that HD~117044 is a binary with two main sequence stars. Even though the uncertainty on the binary mass function is large compared to our other three targets, we can characterize the binary and reject this as a black hole candidate with a single-epoch SHARK-VIS observation. Additional RVs could be used to better constrain the RV orbit and estimate the masses and inclinations (as in Figure \ref{fig:panel_HD137909}c). The flux ratio is the closest to unity in our sample (Table \ref{tab:sharkvis_results}), but the companion would likely still be challenging to detect spectroscopically without techniques like spectral disentangling \citep[e.g.,][]{Seeburger24}. 

\subsection{HD 176695} \label{sec:HD176695}

\begin{figure}
    \centering
    \includegraphics[width=\linewidth]{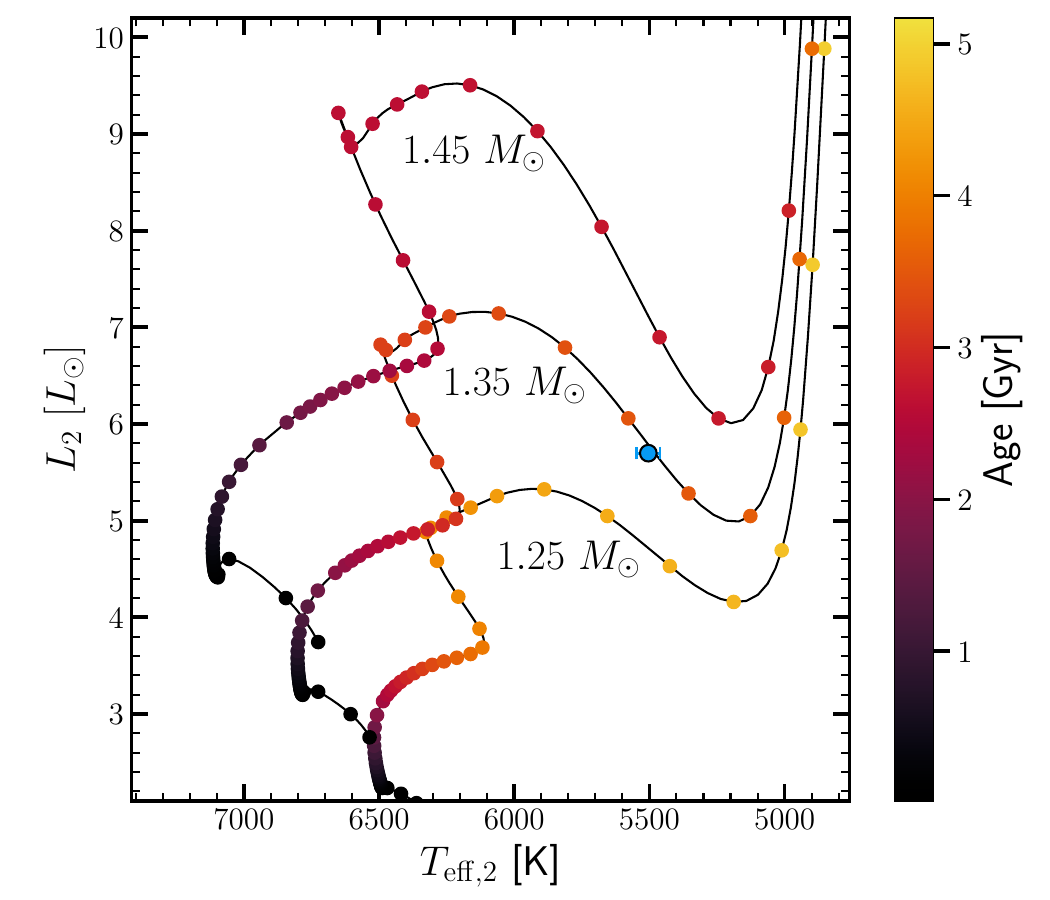}
    \caption{MIST evolutionary tracks for Solar metallicity stars at $1.25$, $1.35$, and $1.45\ M_\odot$ colored by stellar age. The blue point shows the effective temperature and luminosity derived for the secondary of HD 176695 using the two-star SED fit with a prior on the SHARK-VIS flux ratio (Figure \ref{fig:panel_HD176695}d). The $T_{\rm{eff,2}}$ and $L_2$ indicate $1.35\ M_\odot$ companion, but this is not consistent with the RV orbit, so the companion is instead most likely a binary (Figure \ref{fig:HD176695_sed_triple}).}
    \label{fig:HD176695_evo_tracks}
\end{figure}

\begin{figure*}
    \centering
    \includegraphics[width=\linewidth]{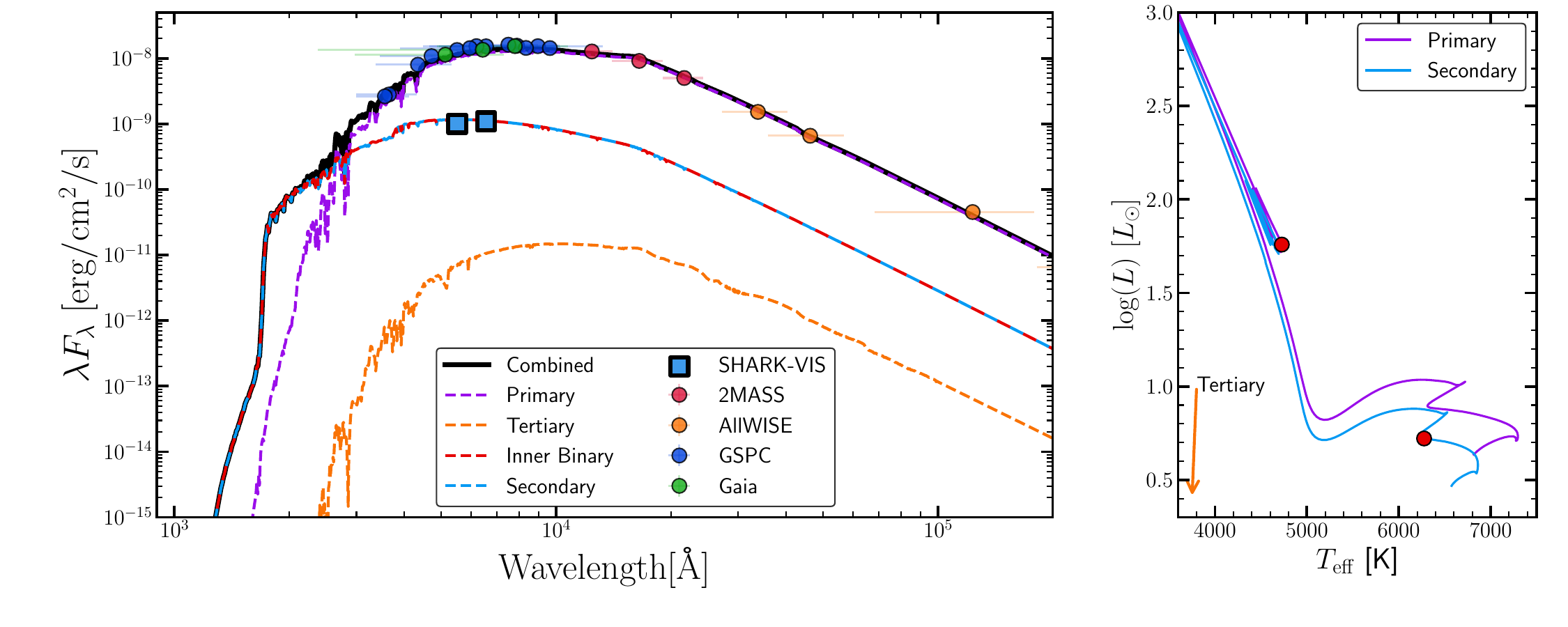}
    \caption{Left: Spectral energy distribution of HD 176695 compared to a hierarchical triple-star SED model. We use MIST evolutionary tracks to explore triple systems with different component masses and search for systems that match the SED, SHARK-VIS flux ratio, and RV orbit constraints. In this example, the primary star is $1.5\ M_\odot$, and the inner binary consists of a $1.37\ M_\odot$ and a $0.56\ M_\odot$ star. The colored lines show the contributions of the various components to the inner binary and the combined SED, and the blue square shows the expected flux ratio between the photometric primary and the inner binary based on the SHARK-VIS observation. Right: evolutionary tracks for the same system. The primary is an evolved red giant and the more massive component of the inner binary is near the end of its main sequence lifetime.}
    \label{fig:HD176695_sed_triple}
\end{figure*}

\renewcommand{\yposdd}{53}
\customfigureilocator{HD176695}{Panel (a): SHARK-VIS $R$-band image of HD 176695. The companion is visible after selecting the best 3\% of frames with a separation of 42.54~mas and an R-band flux ratio $\alpha_R = 0.0914$ (Table \ref{tab:sharkvis_results}) The bright component to the lower left is an artifact of trefoil abberation (Section \S\ref{sec:sharkvis_observations}). Panel (b): RV orbit of HD 176695 based on \citet{Heard56} and \citet{Griffin95} RV measurements The vertical line marks the time of the SHARK-VIS observation. Panel (c): Constraints on the companion mass for different combinations of $M_1$ and $i$ from the RV orbit (black) and SHARK-VIS angular separation (red). The two intersect for inclinations $65^{\circ}$ -- $75^{\circ}$, which corresponds to companion masses between $1.8$--$2.0\ M_\odot$. Panel (d): two-star fit to the multi-band photometry with a prior on the $R$-band flux ratio from SHARK-VIS. Figure \ref{fig:HD176695_evo_tracks} shows the MIST evolutionary tracks, and we find that this solution is not compatible with the RV orbit. It is more likely that the HD~176695 is a triple system, and Figure \ref{fig:HD176695_sed_triple} shows an example of a possible SED for a triple system model. Panel (e): Lili spectrum of the photometric primary (blue) and BT-Settl model atmosphere (black). The giant has strong Paschen lines and some metal lines between 1000--1100~nm but is otherwise relatively featureless.}
\renewcommand{\yposdd}{12}

HD 176695 is a red giant star at $d=290$~pc \citep{BailerJones21}. \citet{Heard56} measured five RVs of this target between 1948 and 1953 from the DDO and identified it as being RV variable. \citet{Griffin95} later took 161 more observations between 1973 and 1995 with CORAVEL, confirming the RV variability and constraining the RV orbit. Figure \ref{fig:panel_HD176695}b shows the combined set of RV measurements and the joint fit to the data. The orbital period is $7710^{+40}_{-50}$~days (21.1~years) and the velocity semi-amplitude is $11.6$~km/s, and this solution agrees with the orbit reported in the SB9 catalog. Despite the CORAVEL observations only covering one orbit, the orbit is well constrained because of the large number of measurements and the dense phase coverage.

The StarHorse2 catalog reports a primary mass of $M_1 = 1.4^{+0.4}_{-0.3}\ M_\odot$, and no other mass estimates are available. Figure \ref{fig:panel_HD176695}c shows the constraints on the companion mass as a function of the orbital inclination. Even if the mass of the primary is lower than predicted, $M_1=1.1\ M_\odot$ and the orbit is edge-on, the companion must be $>1.54\ M_\odot$. If $M_2 > M_1$, the companion would also have to be a giant of greater or comparable luminosity, yet no companion is observed in the spectrum \citep{Griffin95}. The mass ratio can only be $q \leq 1$ if the primary is more massive, $\gtrsim 2.1\ M_\odot$ and the orbit is edge-on, which would mean the secondary is an A/F star. Based on the RV orbit and primary mass estimate, the maximum angular separation is 38~mas for an edge-on orbit.

We observed HD 176695 with SHARK-VIS on 2024-05-17 in $R$- and $V$-band using a 3ms integration time for a total exposure time of 300s with a plate scale of 6.43~mas/pix for both filters. Figure \ref{fig:panel_HD176695}a shows that the companion is visible in the registered and sky-oriented average stack built from the best 3\% (47\%) of frames in the $R$-band ($V$-band). We use our Bayesian Modeling approach (Section \ref{sec:HD137909}) to determine a companion $R$-band flux ratio of $(9.14\pm0.02)\times10^{-2}$ and a separation of $42.54\pm0.02$~mas (Table \ref{tab:sharkvis_results}). The $V$-band flux ratio is $(10.0\pm0.6)\times10^{-2}$. As with HD 117044, there is trefoil abberation around the primary in the SHARK-VIS image. We can distinguish this instrumental artifact from the astrophysical companion because the abberation is fixed with respect to the telescope and has the same orientation in all frames before de-rotation.

Figure \ref{fig:panel_HD176695}c shows the companion mass as a function of orbital inclination from the RV orbit and projected separation constraints for different estimates of the primary mass. The measured angular separation ($42.54\pm0.02$~mas) requires inclinations of $\sim 65^{\circ}$ -- $75^{\circ}$ and companion masses between $1.8$--$2.0\ M_\odot$ for a $1.1$--$1.8\ M_\odot$ primary. An additional SHARK-VIS observation could measure $M_1$ as well (See Section \S\ref{sec:HD137909}).

We fit the \Gaia{}, 2MASS, and \WISE{} photometry with a two-star SED model with a prior on the $R$- and $V$-band flux ratios from SHARK-VIS. There is no previous spectroscopic metallicity, so we assume $[\rm{Fe/H}]=0.0$ for the two-star SED fit. Figure \ref{fig:panel_HD176695}d shows the two-star SED fit to HD~176695. The MCMC model finds a total luminosity of $66\ L_\odot$ where the companion has $T_{\rm{eff,2}}=5500$~K and luminosity $L_2=5.7\ L_\odot$. For MIST evolutionary tracks \citep{Dotter16,Choi16} of Solar-metallicity stars of different masses we find that this combination of effective temperature and luminosity is consistent with a $\sim 1.35\ M_\odot$ companion (Figure \ref{fig:HD176695_evo_tracks}). 

However, a companion mass $M_2 = 1.35\ M_\odot$ is not consistent with the observed RV orbit and angular separation. From the RV orbit constraint alone, if the companion was $1.35\ M_\odot$, the primary would have to be $\gtrsim 0.85\ M_\odot$. Such a system can be rejected because the companion would be more evolved than the primary and the total luminosity would be much less than the $66.0\ L_\odot$ needed to match the observed photometry. A system with $M_2 = 1.35\ M_\odot$ would also not match the combined projected angular separation, even as $M_1$ goes to zero.

This means that the most likely companion to HD 176695 is an inner binary. With only optical+IR photometry and the flux ratio in two optical bands, we cannot fit for the parameters of the inner binary components. We instead combine MIST evolutionary tracks and \citet{Castelli03} model atmospheres to search for systems that match the constraints from the RV orbit, flux ratio, projected separation, and combined SED. Figure \ref{fig:HD176695_sed_triple} shows an example system with $M_1=1.5\ M_\odot$, $M_2=1.37\ M_\odot$, and $M_3 = 0.56\ M_\odot$. The SED is dominated by the evolved primary, and $M_2$ is near the end of its main sequence lifetime, contributing the $9\%$ to the $R$-band flux necessary to match the SHARK-VIS flux ratios. The low-mass star in the inner binary contributes negligibly to the observed flux, but adds enough mass to make $M_2 + M_3 > M_1$, satisfying the RV orbit and observed separation constraints. We emphasize that this is simply a possible solution. More observations are needed to characterize the inner binary. Ultraviolet observations could be used to search for evidence of the inner binary component with $M_2 \sim 1.37\ M_\odot$ since it dominates the flux at these wavelengths. Unfortunately, since HD~176695 is near the Galactic plane ($b=10.9^{\circ}$), it is not included in the \GALEX{} survey and no other archival UV observations are available. 

The tertiary in this proposed model ($M_3=0.56\ M_\odot$) is most likely a main sequence star rather than a white dwarf. This is because the progenitor of a $0.56\ M_\odot$ WD would have a zero-age main sequence mass of $\sim 1.1\ M_\odot$ based on the IFMR \citep{Hollands24}. Such a progenitor would not have formed a WD before the $M_1=1.5\ M_\odot$ and $M_2=1.37\ M_\odot$ stars. However, this triple system model is sensitive to the assumed primary mass, and additional SHARK-VIS observations could be used to determine $M_1$ and measure $M_2+M_3$.

Finally, we also obtained a spectrum of the photometric primary using Lili. As with HD~104438, the companion was not identified in the acquisition camera and we did not measure a spectrum of the companion. Figure \ref{fig:panel_HD176695}e shows the Lili spectrum of the red giant and a model atmosphere for comparison. The spectrum is relatively featureless, aside from the Paschen lines. 

In summary, HD~176695 is most likely a hierarchical triple. Based on the RV orbit, the companion must be more massive than red giant primary unless the giant is massive $>2\ M_\odot$. However, such a binary is not consistent with the observed SED or the SHARK-VIS flux ratio. The only way to construct a system that matches the SED, RV, and SHARK-VIS flux ratio constraints is have the companion be a binary with an unequal mass ratio (Figure \ref{fig:HD176695_sed_triple}).

\section{Discussion and Conclusions} \label{sec:discussion}

We observed four bright ($R\lesssim 8$~mag) nearby ($d\lesssim 300$~pc) SB1s included in the SB9 catalog of spectroscopic binaries. We selected these targets based on their mass functions and projected angular separations (Figure \ref{fig:selection_figure}) as candidates for hosting massive, dark companions. We used SHARK-VIS to directly resolve these systems and look for evidence of a second luminous star in the system and use Lili for spectroscopic characterization.

For all four systems, we identify a luminous companion and rule out a BH/NS for three systems. We model the archival RV observations and the multi-band photometry to characterize the companions. Only two of the four systems, HD~137909 and HD~117044, are consistent with simple stellar binaries. We find that HD~104438 and HD~176695 are likely triples. 

HD~137909 (Section \S\ref{sec:HD137909}) is a massive main sequence binary. The companion was previously identified with the VLT CONICA infrared observations \citep{Bruntt10}. We measure a SHARK-VIS flux ratio of $0.182\pm0.001$, which is consistent with the \citet{Bruntt10} SED model, where the companion is another slightly less massive main sequence star (Figure \ref{fig:panel_HD137909}d). By combining the orbital separation measurement from VLT/NACO and SHARK-VIS, we measure the orbital inclination and both component masses (Figure \ref{fig:panel_HD137909}c).

HD~104438 (Section \S\ref{sec:HD104438}) is red clump star and no companion has been previously detected. The RV orbit is long, $P=35.6$~yr, and we identify a faint companion separated by $107.3$~mas. Based on the RV orbit, this star is not the source of the RV variability and is likely a wide tertiary (Figure \ref{fig:panel_HD104438}c). We find that it is unlikely to be a chance alignment with a background star. The small flux ratio $(1.90\pm 0.02)\times10^{-4}$ (Table \ref{tab:sharkvis_results}) implies that it is most likely an M-dwarf (Figure \ref{fig:panel_HD104438}). Since the binary companion to the red giant is not detected, we cannot rule out a compact object companion. Additional RV observations are needed to refine the ephemeris and rule out a luminous star. Since the orbit is now past periastron and the separation is increasing (Figure \ref{fig:HD104438_angular_sep}), additional SHARK-VIS observations could be used to search for a luminous companion. 

HD~117044 (Section \S\ref{sec:HD117044}) is a main sequence star, and the RV orbit is poorly constrained so the mass function is uncertain. We identify a luminous companion in the SHARK-VIS image separated by $30.42\pm0.50$~mas (Figure \ref{fig:panel_HD117044}), and it is consistent with a second, lower mass main sequence star, suggesting that the mass function is low, $f(M)\sim 0.3\ M_\odot$. The binary has the smallest projected angular separation in our sample, but the companion is clearly detectable in the SHARK-VIS image even by eye.

Finally, HD~176695 (Section \S\ref{sec:HD176695}) is an evolved giant with a large mass function, $f(M)=0.52\ M_\odot$. Figure \ref{fig:panel_HD176695}c shows that the combination of the mass funciton and the measured angular separation requires $M_2 > M_1$. The companion should then also be a red giant, but the SHARK-VIS observations reveal a companion with an $R$-band flux ratio of $(9.14\pm0.02)\times10^{-2}$ (Table \ref{tab:sharkvis_results}). While we can find a solution by fitting the SED with a two-star SED model using a prior on the $R$- and $V$-band flux ratios, comparisons to MIST evolutionary tracks show that such a system is not consistent with the RV orbit (Figure \ref{fig:HD176695_evo_tracks}). Instead, we find that it is more likely that HD~176695 is a triple system, with an unequal mass inner binary made up of an intermediate mass and low mass star. Figure \ref{fig:HD176695_sed_triple} shows an example of the SED fit and evolutionary tracks for an example system that meets the constraints of the RV orbit, combined SED, and SHARK-VIS flux ratio. Additional UV photometry and flux ratio measurements at other wavelengths are needed to confirm and characterize the inner binary. 

High-contrast imaging is typically used to search for and characterize exoplanets \citep[e.g.,][]{Chauvin04, Marois08}. The planetary companions are orders of magnitude fainter than the host stars, and techniques such as ADI are needed to carefully model and remove the signal of the host star. The observations of the binary targets described here are much more straightforward. We expect the luminous stellar companions to have a flux ratio of $\gtrsim 1\%$ for evolved photometric primaries and $\gtrsim 10\%$ for main sequence primaries. For three out of four systems, the companions are easily identifiable in the stacked image or after a best frames selection. PCA-ADI was only necessary for identifying the likely tertiary companion to HD 104438 (Figure \ref{fig:panel_HD104438}). While it is possible that luminous companions to these system could have been identified with multiple high-resolution spectra using techniques like spectral disentangling \citep[e.g.,][]{Ilijic04, Seeburger24}, here we use only a single SHARK-VIS observation with a typical total time on-source of $\sim 10$~min to directly detect the luminous companion. This is especially useful for long period systems or systems with uncertain mass functions, like HD~117044, where multiple years of RV observations would likely be needed to confirm the orbit and search for a black hole companion. Only one SHARK-VIS observation is needed to confidently reject this system as a non-interacting black hole candidate for this system. Since the observation time is short, future SHARK-VIS observations will be done in multiple filters/bands to better constrain the companion SEDs. 

Combining astrometric and spectroscopic orbits to constrain binary orbits is a well-established technique \citep[e.g.,][]{Morbey75}. High-contrast imaging has also been combined with RVs to measure the masses of planets, brown dwarfs, and stars \citep[e.g.,][]{Crepp12, Boehle19, Brandt19}. Although the primary objective of our survey is to identify stars with compact object companions, the systems presented here show how SHARK-VIS can be used to make benchmark stellar mass measurements. Combining the SHARK-VIS separation with the radial velocity mass function determines the companion mass $M_2$ and orbital inclination $i$ given the mass of the primary. With a second measurement of the projected orbital separation, as was available from the earlier VLT/NACO observations of HD~137909 (Section \S\ref{sec:HD137909}), the primary mass can be directly determined as well. This demonstrates that it would be straightforward to use SHARK-VIS to fully characterize the properties of the $\gtrsim 300$ systems with separations larger than $\sim 30$~mas in Figure \ref{fig:selection_figure} using two SHARK-VIS observations.

Two of the four systems we observed are most likely triple systems. Triples have been previously suggested as false positives in the search for non-interacting black holes and neutron stars \citep[e.g.,][]{vandenHeuvel20}, but by far the most common false positives have been stripped stars \citep[e.g.,][]{Jayasinghe22} and massive white dwarfs \citep[e.g.,][]{Tucker24}. Direct imaging surveys are less likely to find stripped stars, since post-mass transfer binaries will in general have shorter orbital separations than those we can resolve with SHARK-VIS or similar instruments. Massive white dwarfs could be detected through this method, and multiple observations could be used to place constraints on the orbital inclination and measure the companion mass more precisely.

These four systems highlight how SHARK-VIS can efficiently search for stars with compact object companions. One of our targets remains a candidate for hosting a compact object, and we show how two measurements of the projected separation can break the degeneracy between $M_1$ and $i$. There are $>100$ other candidates in the SB9 and \Gaia{} SB1 catalogs (Figure \ref{fig:selection_figure}) that can be observed by SHARK-VIS. Instruments with smaller angular resolution, such as VLTI GRAVITY \citep{Gravity17}, could be used to expand the search to binaries with separations $< 10$~mas. Future \Gaia{} data releases are also expected to expand the sample of binaries with astrometric and spectroscopic orbits. Some of these will likely also be candidates for direct imaging where a luminous companion can be identified with a single observation.

\section*{Acknowledgements}

We thank Kris Stanek, Michael Tucker, Ji Wang, Jennifer Rodriguez, and Paarmita Pandey for helpful discussions. DMR is supported by the OSU Presidential Fellowship. CSK is supported by NSF grants AST-1907570, 2307385, and 2407206.

J. Roth acknowledges financial support from the German Federal Ministry of Education and Research (BMBF) under grant 05A23WO1 (Verbundprojekt D-MeerKAT III).

Observations obtained using iLocater are based upon work supported by the National Science Foundation under Grant No. 1654125 and 2108603.

The LBT is an international collaboration among institutions in the United States, Italy, and Germany. LBT Corporation partners are: The University of Arizona on behalf of the Arizona Board of Regents; Istituto Nazionale di Astrofisica, Italy; LBT Beteiligungsgesellschaft, Germany, representing the Max-Planck Society, The Leibniz Institute for Astrophysics Potsdam, and Heidelberg University; The Ohio State University, representing OSU, University of Notre Dame, University of Minnesota, and University of Virginia. PEPSI was made possible by funding through the State of Brandenburg (MWFK) and the German Federal Ministry of Education and Research (BMBF) through their Verbundforschung grants 05AL2BA1/3 and 05A08BAC. 

This work presents results from the European Space Agency space mission Gaia. Gaia data are being processed by the Gaia Data Processing and Analysis Consortium (DPAC). Funding for the DPAC is provided by national institutions, in particular the institutions participating in the Gaia MultiLateral Agreement.

\clearpage
\bibliography{sharkvis}{}
\bibliographystyle{aasjournal}



\end{document}